\documentclass[prd,aps,10pt,twocolumn,nofootinbib,eqsecnum,showpacs]{revtex4-1}

\pdfoutput=1
\usepackage{amssymb,amsmath}
\usepackage{epsfig}
\usepackage{xcolor}
\usepackage[utf8]{inputenc}
\usepackage{stmaryrd}
\usepackage{mathrsfs}

\DeclareMathOperator\erf{erf}

\DeclareMathOperator\erfi{erfi}
\DeclareMathOperator\sgn{sgn}

\newcommand{\W}{\mathscr{W}}

\newcommand{\be}{\begin{equation}}
\newcommand{\ee}{\end{equation}}
\newcommand{\ba}{\begin{eqnarray}}
\newcommand{\ea}{\end{eqnarray}}
\newcommand{\beg}{\begin{gather*}}
\newcommand{\eng}{\end{gather*}}
\newcommand{\hh}{,\hspace{0.5cm}}

\newcommand{\eq}[1]{(\ref{#1})}

\newcommand{\n}[1]{\label{#1}}
\newcommand{\bs}[1]{{\boldsymbol{#1}}}

\newcommand{\ins}[1]{{\mbox{\tiny #1}}}

\def\XXint#1#2#3{{\setbox0=\hbox{$#1{#2#3}{\int}$ }
\vcenter{\hbox{$#2#3$ }}\kern-.6\wd0}}

\usepackage[makeroom]{cancel}
\usepackage[caption=false]{subfig}
\usepackage[colorlinks=true,
            citecolor=green,
            linkcolor=red,
            filecolor=cyan,
            urlcolor=magenta,
            backref=false]{hyperref}

\newcommand{\dd}{\mbox{d}}

\begin{document}

\title{Superradiance in a ghost-free scalar theory}

\author{Valeri P. Frolov}
\email{vfrolov@ualberta.ca}
\affiliation{Theoretical Physics Institute, University of Alberta, Edmonton, Alberta, Canada T6G 2E1}
\author{Andrei Zelnikov}
\email{zelnikov@ualberta.ca}
\affiliation{Theoretical Physics Institute, University of Alberta, Edmonton, Alberta, Canada T6G 2E1}


\begin{abstract}
We study superradiance effect in the ghost-free theory. We consider a scattering  of a ghost-free scalar massless field on a rotating cylinder. We assume that  cylinder is thin  and empty inside, so that its interaction with the field is described by a delta-like potential. This potential besides the real factor, describing its height, contains also imaginary part, responsible for the absorption of the field. By calculating the scattering amplitude we obtained the amplification coefficient both in the local and non-local (ghost-free) models and demonstrated that in the both cases it is greater than 1 when the standard superradiance condition is satisfied. We also demonstrated that dependence of the amplification coefficient on the frequency of the scalar field wave may be essentially modified in the non-local case.
\end{abstract}

\maketitle


\section{Introduction}

In this paper we discuss the superradiance effect in a ghost-free theoriy. In general, the superradiance is a phenomenon where radiation is enhanced. This effect is known in different areas of physics, such as quantum optics, electromagnetism, fluid dynamics and quantum mechanics. The famous example of the superradiance is an amplification of an electromagnetic wave by a rapidly rotating conducting body \cite{zeldovich:1971,zeldovich:1972,Bolotovskii:1975}. If the angular velocity of the rotating body is $\Omega$, then an infalling wave  with the positive frequency $\omega$ and an azimuthal angular momentum $m$ is amplified when the superradiance condition is met
\be \n{SR}
0<\omega<m\Omega .
\ee
Suppose that a rotating body is a cylinder of radius $R$, then the condition \eq{SR} implies that the linear velocity $R\Omega$ of the surface of the cylinder is faster than the phase velocity of the wave. Zel'dovich realized that in the quantum physics there should exist a similar effect of spontaneous emission from vacuum of quanta satisfying the superradiance condition \eq{SR}. He also suggested that the superradiance may exist in rotating black holes. The effect of amplification of the waves by rotating black holes was studied in \cite{Starobinsky:1973aij,Starobinsky:1974nkd,Breuer:1973uc}, while Unruh \cite{Unruh:1974bw} demonstrated the existence of the quantum spontaneous superradiance emission by direct calculations.

The superradiance effect is quite similar to a wide class of phenomena connected with a so-called anomalous Doppler effect \cite{Ginzburg:1993}. For example, an electrically neutral body with internal degrees of freedom, moving uniformly through the media may emit electromagnetic waves even if it starts off in its ground state. It happens when the velocity of the body is higher than the speed of light in the media \cite{Bekenstein:1998nt,Frolov:1986,Ginzburg:1987}.

Recently, the existence of classical effect of the superradiance was demonstrated by studying the behavior of sound and surface waves
\cite{Torres:2016iee,Cardoso:2016zvz}. More recent discussion of the superradiance effect and its interesting astrophysical applications can be found e.g in
\cite{Richartz:2009mi,Vicente:2018mxl,Brito:2015oca}. See also references therein.

As we already mentioned, our purpose is to study the superradiance effect in the framework of the ghost-free theory. In such a theory the field equations are modified by introducing a non-local form factor. The latter is chosen so that it does not introduce ghosts and the number of degrees of freedom of the modified theory is the same as for a local one. In a flat spacetime the corresponding form-factor is Lorentz invariant and usually it has the form $\sim \exp((-\ell^2 \Box)^N)$, where $N$ is a positive integer number and $\ell$ is the characteristic scale where non-local modification becomes important. Ghost-free theories of this type are usually called $GF_N$ theories \cite{Frolov:2016xhq}. Nice recent reviews of ghost-free theories can be found in \cite{Buoninfante:2018mre,Buoninfante:2018xif,Modesto:2017sdr}.
One of interesting applications of ghost-free theories is study of ghost-free modifications of the Einstein gravity \cite{Li:2015bqa,Modesto:2017hzl,Modesto:2017ycz,Cornell:2017irh}. It was demonstrated that such modifications may help to resolve problem of singularities in cosmology and black holes
\cite{Frolov:2015bta,Frolov:2015usa,Frolov:2015bia,Koshelev:2017bxd,Calcagni:2017sov}. In the present paper we restrict ourselves by studying of the effect of superradiance for the case of scalar fields. However the obtained results can be easily generalized to the scattering of ghost-free fields with non-zero spins, including linearized ghost-free gravity.

For study of the superradiance in the ghost-free theory we consider a simple model. Namely we consider a scattering of the ghost-free massless scalar field $\varphi$ on a rotating infinitely long thin cylinder of radius $R$. This interaction is described by $\delta$-like potential, which besides a parameter, characterising its height, contains also some absorption coefficient\footnote{A similar model in the local quantum mechanics was briefly discussed in \cite{Brito:2015oca}}. We describe this model in Section \ref{section2}. We calculate the amplification coefficient in the local theory by two methods giving the same results: by using jump conditions and by solving Lippmann-Schwinger equation (Sections  \ref{section2} and  \ref{section3}). In Section  \ref{section4} we demonstrate that a similar problem is exactly solvable in the ghost-free theory of the scalar massless field. In the Sections \ref{section5}-\ref{section6} we study the properties of the amplification coefficients for the ghost-free case. Section  \ref{section7} contains brief discussion of the obtained results.


\section{Superradiance in a local scalar theory}\label{traditional}\label{section2}

Let us remind a derivation of a classical superradiance effect in a local scalar theory. In the presence of an absorbing medium the massless scalar field satisfies the equation (see Appendix \ref{absorption})\footnote{Equations of this type naturally appear in consideration of superradiance in stars \cite{zeldovich:1971,Cardoso:2015zqa} and many other applications \cite{Barbier:2015,Garner:2017}.}
\be\label{varphi}
\Box\varphi -V\,\varphi=0\, .
\ee
We consider a special case when the operator $V$ has the form
\be\label{V}
V={\delta(\rho-R)\over R}\big(\beta+\alpha R \,u^{\mu}\partial_{\mu}\Big)\, .
\ee
Such a potential describes the matter distribution localized on the surface of a cylinder of radius $R$. Here $u^{\mu}$ is the four-velocity vector of an element of an absorbing medium. The term
proportional to $\beta$ characterizes the "height" of a semi-transparent cylindrical barrier. The other term, which is proportional to $\alpha$, describes an interaction with absorbing medium, which is located on the surface of the rotating cylinder. We chose normalization of the height $\beta$ of the potential and the absorption coefficient $\alpha$ so that they are dimensionless quantities.

The metric in cylindrical coordinates reads
\be
ds^2=-dt^2+dz^2+d\rho^2+\rho^2d\phi^2.
\ee
The linear velocity of the surface of the rotating cylinder is
\be
u^{\mu}={1\over \sqrt{1-{\Omega^2 R^2}}} \big[1,0,0,\Omega\big],
\ee
where $\Omega$ is its angular velocity. This linear velocity tends to the speed of light in the limit $R\Omega\to 1$.

Now we expand the scalar field in modes
\be\label{modes}
\varphi=\sum_{\omega, k,m} e^{-i\omega t+ikz+im\phi}\varphi_{\omega k m}(\rho).
\ee
Here
\be
\sum_{\omega, k,m} \equiv  {1\over2\pi}\sum_{m=-\infty}^\infty\int_{-\infty}^\infty {\dd k\over 2\pi} \int_{-\infty}^\infty {\dd \omega\over 2\pi}.
\ee
For a real field $\varphi$ the radial harmonics obey the property
\be\label{CC}
\varphi_{-\omega -k -m}(\rho)=\varphi^*_{\omega k m}(\rho).
\ee

Every mode $\psi(\rho)\equiv \varphi_{\omega k m}(\rho)$ satisfies a one-dimensional equation
\be\label{psi}
(\hat{F}-V_{\omega m})\psi=0.
\ee
Here
\be\label{hatF}
\hat{F}={1\over \rho}\partial_{\rho}(\rho\,\partial_\rho)+\Big[\omega^2-k^2-{m^2\over \rho^2}]
\ee
and the $\delta$-like complex potential $V_{\omega m}$ is\footnote{Discussion of the properties of a $\delta$-like complex potentials and its application to description of absorbtion dynamics can be found in \cite{Villavicencio:2018}.}
\be\begin{split}\label{V}
&V_{\omega m}=\Lambda\,{\delta(\rho-R)\over R},\\
&\Lambda=\beta+i\gamma,\\
&\gamma=\alpha {(m\Omega-\omega)R\over \sqrt{1-{\Omega^2 R^2}}}.
\end{split}\ee
$V_{\omega m}$ is the Fourier transform of $V$.
Note that parameters $\alpha,\beta,\gamma,\Lambda$ are dimensionless in the units when the light speed $c=1$. The absorption condition on the cylinder corresponds to $\alpha>0$.

For the waves that propagate to infinity ($\rho\to\infty$) there is a condition $\omega^2-k^2> 0$. The solution of \eq{psi} is of the form
\be
\psi=\begin{cases} C_0 J_m(\varpi\rho); & \rho<R,\\
C_1 J_m(\varpi\rho)+C_2 Y_m(\varpi\rho) ; & \rho>R,
\end{cases}
\ee
\be
\varpi=\sqrt{\omega^2-k^2}.
\ee

The continuity and the jump conditions lead to the equations for the complex constants $C_{0,1,2}$
\be
(C_0-C_1) J_m-C_2 Y_m=0,
\ee
\be
\varpi\big[(C_0-C_1) J_{m+1}-C_2 Y_{m+1}\big]={\Lambda\over R} C_0 J_m.
\ee
Here and later, when the argument of the Bessel functions is $\varpi R$, we omit it and denote  $J_m=J_m(\varpi R)$, $Y_m=Y_m(\varpi R)$, $H^{(1)}_m=H^{(1)}_m(\varpi R)$, $H^{(2)}_m=H^{(2)}_m(\varpi R)$.

The solution of these equations reads
\ba
{C_1\over C_0}=1+{\Lambda\over\varpi R}{Y_m\over \Delta}\hh
{C_2\over C_0}=-{\Lambda\over\varpi R}{J_m\over \Delta}\, ,\\
\Delta={J_mY_{m+1}-J_{m+1} Y_m\over J_m}=-{2\over\pi \varpi R} {1\over J_m}\, .
\ea
In the last equality we used the following property of the Wronskian
\be\begin{split}
&\W\{J_m(z),Y_m(z)\}=J_m(z) Y'_m(z)-Y_m(z) J'_m(z)\\
&=J_{m+1}(z)Y_m(z)-Y_{m+1}(z) J_m(z)={2\over \pi z}.\\
\end{split}
\ee

Thus we have
\be
{C_1\over C_0}=1-{\pi\over 2}{\Lambda}J_m Y_m
\hh
{C_2\over C_0}={\pi\over 2}{\Lambda}(J_m)^2.
\ee

At large $\rho\gg \varpi^{-1}$, $\rho > R$ the asymptotic behavior of the solution is
\be\begin{split}
\psi\simeq{1\over\sqrt{2\pi\varpi\rho}}\Big[&(C_1-iC_2)e^{+i\big[\varpi\rho-\pi\big({m\over 2}+{1\over 4}\big)\big]}\\
+&(C_1+iC_2)e^{-i\big[\varpi\rho-\pi\big({m\over 2}+{1\over 4}\big)\big]}\Big].
\end{split}\ee
The outgoing wave corresponds to
\be
\psi\sim \begin{cases}
e^{+i\varpi\rho}& \omega>0,\\
e^{-i\varpi\rho}& \omega<0.
\end{cases}\ee
Correspondingly one can write for $\omega>0$
\be\begin{split}
\psi\simeq{(C_1+iC_2)\over\sqrt{2\pi\varpi\rho}}\Big[A\,& e^{+i\big[\varpi\rho-\pi\big({m\over 2}+{1\over 4}\big)\big]}\\
+&e^{-i\big[\varpi\rho-\pi\big({m\over 2}+{1\over 4}\big)\big]}\Big],
\end{split}\ee
and for $\omega<0$
\be\begin{split}
\psi\simeq{(C_1-iC_2)\over\sqrt{2\pi\varpi\rho}}\Big[& e^{+i\big[\varpi\rho-\pi\big({m\over 2}+{1\over 4}\big)\big]}\\
+\widetilde{A}\,&e^{-i\big[\varpi\rho-\pi\big({m\over 2}+{1\over 4}\big)\big]}\Big],
\end{split}\ee

Here the complex relative amplitudes are
\be\begin{split}\label{A}
A&\equiv {C_1-iC_2\over C_1+iC_2}
={1-i{\pi\over 2}\Lambda J_m H^{(2)}_m \over 1+i{\pi\over 2}\Lambda J_m H^{(1)}_m }\\
&={{2\over\pi  J_m}+\big(\gamma J_m-\beta Y_m\big)-i\big(\beta J_m+\gamma Y_m\big)\over {2\over\pi  J_m}-\big(\gamma J_m+\beta Y_m\big)+i\big(\beta J_m-\gamma Y_m\big)}
\end{split}\ee
and
\be\label{tildeA}
\widetilde{A}\equiv {C_1+iC_2\over C_1-iC_2}={1\over A}.
\ee

The amplification factor $Z$ is defined as
\be
Z=\begin{cases}
|A|^2-1,&\omega>0,\\
|\widetilde{A}|^2-1,&\omega<0.
\end{cases}\ee
Thus we have
\be\begin{split}
|A|^2
&={{\big[{2\over\pi  J_m}+\big(\gamma J_m-\beta Y_m\big)\big]^2+\big(\beta J_m+\gamma Y_m\big)^2}
\over
{\big[{2\over\pi  J_m}-\big(\gamma J_m+\beta Y_m\big)\big]^2+\big(\beta J_m-\gamma Y_m\big)^2}}.
\end{split}\ee
The amplitude $|\widetilde{A}|^2$ can be obtained from $|A|^2$ by substitution $\gamma\to-\gamma$ or
\be
|\widetilde{A}|^2=|A|^{-2}.
\ee
Hence, for positive frequencies we get
\be\begin{split}\label{Z0}
Z&={8\gamma\over\pi }\,{ 1 \over
{\big({2\over\pi  J_m}-\gamma J_m-\beta Y_m\big)^2+\big(\beta J_m-\gamma Y_m\big)^2}}.
\end{split}\ee
Depending on the sign of the parameter $\gamma$ this amplification factor can be either positive or negative.
One can see that superradiance amplification occurs when $Z>0$. It leads to the condition $\gamma>0$, or
\be
0<\omega<m\Omega.
\ee
This condition means that the angular phase velocity of the mode
\be
\Omega_\ins{Phase}={d\phi\over dt}={\omega\over m}
\ee
obeys the inequality
\be
0<\Omega_\ins{Phase}<\Omega.
\ee
In this derivation we assumed $\omega>0$. Because the velocity of the surface of a cylinder can not exceed the speed of light we also have a restriction
$0<R\Omega<1$. It means, in particular, that the argument $\varpi R$ of the Bessel functions entering \eq{Z0} lies in the interval
\be
0\le \varpi R\le m\hh m>0.
\ee
In this range of the argument one has $J_m>0$ and $Y_m<0$. Note that the first zeros $j_{m,1}$ and $y_{m,1}$ of $J_m$ and $Y_m$, correspondingly, obey the inequality (see Eq.(10.21.3) in the book \cite{Olver:2010})
\be
j_{m,1}>y_{m,1}>m.
\ee
Therefore for the case $\beta\ge 0$ the denominator in \eq{Z0} never vanishes and, hence, amplification factor does not diverge.

For negative frequencies one can derive a similar expression for the amplification factor
\be\begin{split}\label{tildeZ0}
\widetilde{Z}&=-{8\gamma\over\pi }\,{ 1 \over
{\big({2\over\pi  J_m}+\gamma J_m-\beta Y_m\big)^2+\big(\beta J_m+\gamma Y_m\big)^2}}.
\end{split}\ee
The amplification factor of the complex conjugated mode $\varphi^*_{\omega k m}(\rho)$ can be obtained from \eq{tildeZ0} by substitution $m\to -m$ and $\gamma\to-\gamma$. As a result it gives exactly the same expression \eq{Z0} as for the $\varphi_{\omega k m}(\rho)$ mode. Thus, every component of a real wave, which is the sum
\be
e^{-i\omega t+ikz+im\phi}\varphi_{\omega k m}(\rho)+e^{i\omega t-ikz-im\phi}\varphi^*_{\omega k m}(\rho),
\ee
has the same amplification factor  \eq{Z0}.

\section{Lippmann–Schwinger method}\label{section3}

Now let us recalculate the amplification factor using the Lippmann–Schwinger approach.
The solution of \eq{psi} can be written in terms of the Green function  $G_0$ of the operator $\hat{F}$
\be\label{psiLS}
\psi(\rho)=\psi_0(\rho)-\int \dd\rho' \rho' \,G_0(\rho,\rho') V(\rho')\psi(\rho'),
\ee
where $\psi_0$ is a solution of the free equation
\be
\hat{F}\psi_0=0.
\ee
The Green function $G_0(\rho,\rho')$ satisfies an inhomogeneous equation
\be
\hat{F}\,G_0(\rho,\rho')=-{\delta(\rho-\rho')\over \rho'}.
\ee
and outgoing asymptotic condition $G_0(\rho,\rho')$
\be
G_0(\rho,\rho')\sim {1\over \sqrt{\rho}}\,e^{i\varpi\rho} \hh  \rho\gg\rho',
\ee
for  $\rho\to \infty$ .
$G_0(\rho,\rho')$ is the Fourier transform of the retarded Green function.
\be
G_0(\rho,\rho')=\int \dd t\dd z\dd\phi\, e^{i\omega t-i kz -i m\phi} G_R(x,x'),
\ee
where
\be
\Box \, G_R(x,x')=-\delta(x,x').
\ee
Substitution of \eq{V} to \eq{psiLS} leads to an algebraic equation
\be\label{psiEQ}
\psi(\rho)=\psi_0(\rho)-\Lambda\,G_0(\rho,R)\psi(R).
\ee
If the complex expression
\be
1+\Lambda G_0(R,R)\neq 0,
\ee
we obtain a solution
\be\label{Lipp}
\psi(\rho)=\psi_0(\rho)-{\Lambda\,G_0(\rho,R)\over 1+\Lambda G_0(R,R)}\psi_0(R).
\ee

The Green function  $G_0(\rho,\rho')$ can be expressed in terms of the Bessel functions
\be\label{G0}
G_0(\rho,\rho')=i {\pi\over 2}\,J_{m}(\varpi\rho_{<}) \times
\begin{cases}~~H^{(1)}_{m}(\varpi\rho_{>}),& \omega>0,\\
-H^{(2)}_{m}(\varpi\rho_{>}),& \omega<0,
\end{cases}
\ee
where
\be
\rho_{<}=\min(\rho,\rho')\hh \rho_{>}=\max(\rho,\rho').
\ee
For $\rho\gg\rho'$ the asymptotic reads
\be\begin{split}
G_0(\rho,\rho')\Big|_{\rho\to\infty}&=i \sqrt{\pi\over 2\varpi\rho}\,J_{m}(\varpi\rho')\\
&\times
\begin{cases}
+e^{+i\varpi\rho-i{\pi\over 2}m-i{\pi\over 4}},&\omega>0,\\
-e^{-i\varpi\rho+i{\pi\over 2}m+i{\pi\over 4}},&\omega<0.
\end{cases}\end{split}
\ee

Now we choose $\psi_0$ describing a regular at the center wave of some arbitrary amplitude. It consists of an incoming and outgoing waves of the same amplitude
\be\label{psi0}
\psi_0=2C\,J_{m}(\varpi\rho)=C[H^{(2)}_{m}(\varpi\rho)+H^{(1)}_{m}(\varpi\rho)].
\ee
Then at large $\rho$ the solution with the potential \eq{Lipp} consists of an incoming and outgoing waves \eq{psi0}
\be
\psi=\psi_\ins{in}+\psi_\ins{out},
\ee
\be
\psi_\ins{in}=C \begin{cases}H^{(2)}_{m}(\varpi\rho),&\omega>0,\\
H^{(1)}_{m}(\varpi\rho),&\omega<0.
\end{cases}
\ee
Using \eq{Lipp} in the case of $\omega>0$ one can derive
\be\begin{split}
&\psi_\ins{out}=C H^{(1)}_{m}(\varpi\rho)-2C{\Lambda\,G_0(\rho,R)\over 1+\Lambda G_0(R,R)}J_{m},\\
&G_0(\rho,R)=i {\pi\over 2}\,J_{m} H^{(1)}_{m}(\varpi\rho),
\\
&G_0(R,R)=i {\pi\over 2}\,J_{m} H^{(1)}_{m},\\
&\psi_\ins{out}=C H^{(1)}_{m}(\varpi\rho){1-i{\pi\over 2}\Lambda J_m H^{(2)}_{m}
\over
1+i{\pi\over 2}\Lambda J_m H^{(1)}_{m}.
}
\end{split}\ee
In the case of $\omega<0$ similar calculations give
\be\begin{split}
&\psi_\ins{out}=C H^{(2)}_{m}(\varpi\rho)-2C{\Lambda\,G_0(\rho,R)\over 1+\Lambda G_0(R,R)}J_{m},\\
&G_0(\rho,R)=-i {\pi\over 2}\,J_{m} H^{(2)}_{m}(\varpi\rho),
\\
&G_0(R,R)=-i {\pi\over 2}\,J_{m} H^{(2)}_{m},
\\
&\psi_\ins{out}=C H^{(2)}_{m}(\varpi\rho){1+i{\pi\over 2}\Lambda J_m H^{(1)}_{m}
\over
1-i{\pi\over 2}\Lambda J_m H^{(2)}_{m}.
}
\end{split}\ee

The complex relative amplitudes $A$ and $\widetilde{A}$ are defined as the ratio of amplitudes of the incoming and outgoing waves evaluated at infinity. Therefore one gets
\be
A={1-i{\pi\over 2}\Lambda J_m H^{(2)}_{m}
\over
1+i{\pi\over 2}\Lambda J_m H^{(1)}_{m}}\hh\widetilde{A}=A^{-1}.
\ee
This result exactly reproduces previously calculated values \eq{A}-\eq{tildeA}.


\section{Superradiance in the Ghost-free scalar theory}\label{section4}

The standard method, as it is described in section \ref{traditional}, is not applicable to the case of the non-local  ghost-free theory. However, the Lippmann–Schwinger approach works pretty well \cite{Boos:2018kir}.

The ghost-free scalar field satisfies the equation
\be\label{GFvarphi}
a(\Box) \Box\varphi -V\phi=0.
\ee
The function $a(z)$ is an analytical function having the form of an exponent of an entire function and is chosen such that $a(0)=1$. The following choice of the form-factor $a$ is often used in the literature
\be
a(\Box)=\exp[ (-\ell^2 \Box)^N]\, .
\ee
We refer for a ghost-free theory of this type as to $GF_N$ theory \cite{Frolov:2016xhq}.

Expanding in modes \eq{modes} one gets the equation for the modes
\be\label{GFpsi}
[a(\hat{F})\hat{F}-V_{\omega m}]\psi=0,
\ee
where the operator $\hat{F}$ and the potential $V_{\omega m}$ are given by \eq{hatF} and \eq{V}.
Let $\psi_0$ be a solution of the homogeneous equation
\be\label{GFpsi00}
\hat{F}\psi_0=0.
\ee
Evidently, it satisfies also the nonlocal equation without a potential
\be\label{GFpsi0}
a(\hat{F})\hat{F}\psi_0=0.
\ee
Because of the properties of the ghost-free theories the form-factor $a(\hat{F})$ does not lead to any extra freely propagating modes.

Let us also introduce the Green function $G_0^\ins{GF}(\rho,\rho')$ as
a solution of the equation
\be
a(\hat{F})\hat{F}\,G_0^\ins{GF}(\rho,\rho')=-{\delta(\rho-\rho')\over \rho'}\, ,
\ee
which is regular at $\rho=0$ and obeys the outgoing boundary conditions.

The solution for \eq{GFpsi} can be found using the Lippmann–Schwinger approach when applied to the non-local operator $a(\hat{F})\hat{F}$. This solution is similar to \eq{Lipp}. The only difference is that $G_0^\ins{GF}$ is substituted for $G_0$
\be\label{GFLipp}
\psi(\rho)=\psi_0(\rho)-{\Lambda\,G_0^\ins{GF}(\rho,R)\over 1+\Lambda G_0^\ins{GF}(R,R)}\psi_0(R).
\ee
It should be emphasized that, though the solutions $\psi_0$ of the {\it homogeneous} equations in the local and non-local cases coincide, the Green functions $G_0^\ins{GF}$ and $G_0$ differ, because they satisfy the {\it inhomogeneous} equations with $\delta$-like source.

The Green function $G_0^\ins{GF}$ can be computed using the momentum representation of the retarded Green function corresponding to the  operator $a(\Box)\Box$
\be\label{GFret}
a(\Box)\Box \, G_R^\ins{GF}(x,x')=-\delta(x,x').
\ee
In the non-local ghost-free theory the standard time ordering does not work well in the vicinity of the null cone and the
retarded Green function  is defined as the solution of \eq{GFret} which satisfies the same asymptotic conditions as the local $G_R(x,x')$ when $|x-x'|^2\to \infty$, that is far away from the null cone. This requirement, in particular, means that
\be
G_0^\ins{GF}(\rho,R)\big|_{(\rho-R)\to\infty}\to G_0(\rho,R).
\ee
Similarly to the local case the amplitude of the reflected wave can be defined using \eq{GFLipp} in the asymptotic domain at large $\rho$. One can see that in this asymptotic the only difference between local and ghost-free  cases is the value of $G_0^\ins{GF}(R,R)$ instead of $G_0(R,R)$ in the denominator of the second term in \eq{GFLipp}.

Using momentum representation in the Cartesian coordinates $X=(t,\bs{x})=(t,x,y,z)$, where $x=\rho\sin\phi$ and $y=\rho\cos\phi$, the retarded Green functions in the local and ghost-free theories are
\be\begin{split}
G_R(X-X')=&-\int{d\omega d \bs{k}\over (2\pi)^{4}}\,e^{-i\omega(t-t')+i \bs{k}( \bs{x}- \bs{x}')}\\
&\times{1\over \nu^2+i\sgn(\omega)\epsilon },
\end{split}\ee
\be\begin{split}
G_R^\ins{GF}(X-X')=&-\int{d\omega d \bs{k}\over (2\pi)^{4}}\,e^{-i\omega(t-t')+i \bs{k}( \bs{x}- \bs{x}')}\\
&\times{a^{-1}(\nu^2)\over \nu^2+i\sgn(\omega)\epsilon }.
\end{split}\ee
Here
\be
\nu^2=\omega^2-\bs{k}^2=\varpi^2-k_x^2-k_y^2\hh
 \bs{k}^2=k_x^2+k_y^2+k_z^2,
\ee
and $\epsilon$ is an infinitesimal positive constant.
Because of  analytical properties of the function $a(\nu^2)$ the retarded Green function, as well as the advanced Green function and the Feynman propagator, can be written in the  form
\be\begin{split}
&G_{R,A,F}^\ins{GF}({X}-{X}')=G_{R,A,F}({X}-{X}')+{\Delta G}({X}-{X}').
\end{split}\ee
Here ${\Delta G}$ is a universal non-local correction to the local propagators
\be\begin{split}\label{DeltaG}
{\Delta G}({X}-{X}')&=\int{d\omega d \bs{k}\over (2\pi)^{4}}\,e^{-i\omega(t-t')+i \bs{k}( \bs{x}- \bs{x}')}\\
&\times{1-a^{-1}(\nu^2)\over \nu^2}.
\end{split}\ee
It is this universal correction ${\Delta G}$, which is responsible for the violation of local causality in the vicinity of the null cone \cite{Buoninfante:2018mre}.
The integrand in this expression does not have poles and, hence, the integral is well defined and unambiguous. Note that in order  to compute this integral one {\it does not} need to analytically continue this expression to the complex plane. Moreover, analytical continuation would be a bad idea, because $a^{-1}(\nu^2)$ may diverge in some complex directions and, hence, the contour integration is not well behaved at infinities.


\section{Superradiance in the $\mathrm{\bf GF}_1$ scalar theory}\label{section5}

Let us consider the $GF_1$ scalar theory \cite{Frolov:2016xhq} which corresponds to the choice
\be\label{a1}
a(\Box)=e^{-\ell^2\Box}.
\ee
Here $\ell$ is the characteristic length scale of non-locality. Typically it is assumed to be larger or of the order of the Planck scale.
Substituting \eq{a1} to \eq{DeltaG} and using an integral representation
\be
{1-e^{\ell^2\nu^2}\over \nu^2}=-\int_0^{\ell^2}ds\,e^{s\nu^2},
\ee
one can write
\be\begin{split}\label{DeltaG1}
{\Delta G}({X}-{X}')&=-\int_0^{\ell^2}ds\,\int{d\omega d \bs{k}\over (2\pi)^{4}}\\
&\times e^{-i\omega(t-t')+i \bs{k}( \bs{x}- \bs{x}')}\,e^{s(\omega^2-\bs{k}^2)}.
\end{split}\ee
Performing the Fourier transform in $t,z$ coordinates first we obtain
\be\begin{split}\label{DeltaG2}
&{\Delta G}_{\omega k}(x-x',y-y')=-\int_0^{\ell^2}ds\,e^{s\varpi^2}\\
&\int_{-\infty}^{\infty}{dk_x dk_y\over (2\pi)^{2}} e^{ik_x(x-x')+ik_y(y-y')}\,e^{-s (k_x^2+k_y^2)},
\end{split}\ee
where
\be
k=k_z  \hh
\varpi=\sqrt{\omega^2-k^2}.
\ee
Using the coordinate transformation $x=\rho\sin\phi$,  $y=\rho\cos\phi$ we express it in polar coordinates and compute the angular Fourier component
\be\begin{split}\label{DeltaG2}
{\Delta G}_{\omega k m}(\rho,\rho')&=-\int_0^{\ell^2}ds\,e^{s\varpi^2}\int_{-\pi}^{\pi}d\phi \,e^{im(\phi-\phi')}\\
&\int_{-\infty}^{\infty}{dk_x dk_y\over (2\pi)^{2}}\,e^{-s (k_x^2+k_y^2)}\\
&e^{ik_x[\rho\sin\phi-\rho'\sin\phi']}e^{ik_y[\rho\cos\phi-\rho'\cos\phi']}\\
=&-\int_0^{\ell^2}{ds\over 4\pi s}\,\,e^{s\varpi^2}\int_{-\pi}^{\pi}d\phi \,e^{im(\phi-\phi')}\\
& e^{-{\rho^2+\rho'{}^2-2\rho\rho'\cos(\phi-\phi')\over 4 s}}.
\end{split}\ee
The result of integration over the angle variable can be expressed in terms of the modified Bessel function $I_m$
\be\begin{split}\label{DeltaG6}
{\Delta G}_{\omega k m}(\rho,\rho')&=-2\pi\int_0^{\ell^2}{ds\over s}\,e^{s\varpi^2-{\rho^2+\rho'{}^2\over 4 s}}\,I_m\big({\rho\rho'\over 2 s}\big).
\end{split}\ee
One can see that this non-local correction to the propagators is finite, real, and {\it negative} for all values of its real arguments.

The asymptotics  of the Bessel function are
\be\begin{split}
&I_m(x)={1\over \sqrt{2\pi x}}e^{x}\Big [1+O\Big({1\over x}\Big)\Big]\hh x\to \infty,
\\
&I_m(x)={1\over 2^m m !}x^m+ O(x^{m+2})\hh x\to 0.
\end{split}\ee
It means that the asymptotic of the ${\Delta G}_{\omega k m}$, when $\rho\rho'\gg\ell^2$, reads
\be\begin{split}\label{DeltaGass}
{\Delta G}_{\omega k m}(\rho,\rho')&\approx -{2\sqrt{\pi}}{1\over\sqrt{\rho\rho'}}\int_0^{\ell^2}{ds\over \sqrt{s}}\,e^{s\varpi^2-{(\rho-\rho')^2\over 4 s}}.
\end{split}\ee
One can write is in an explicit form
\be\begin{split}\label{DeltaGass1}
{\Delta G}_{\omega k m}(\rho,\rho')&\approx {i\pi\over\varpi\sqrt{\rho\rho'}}\\
\times&\Big\{e^{i\varpi(\rho-\rho')}\Big[\erf\Big({\rho-\rho'\over 2\ell}+i\varpi\ell\Big)-1\Big]\Big\}\\
-&e^{-i\varpi(\rho-\rho')}\Big[\erf\Big({\rho-\rho'\over 2\ell}-i\varpi\ell\Big)-1\Big]\Big\}.
\end{split}\ee

For a fixed $\rho'=R$ and large $\rho\gg R$ the correction to the Green function is exponentially small.
\be\begin{split}\label{Ass1}
{\Delta G}_{\omega k m}(\rho,R)&\sim e^{-{\rho^2\over 2\ell^2}}.
\end{split}\ee
Therefore, it does not affect the asymptotic behavior at large radii.

Computation of the amplification coefficient requires computation of the value
\be\begin{split}\label{g}
g&\equiv{\Delta G}_{\omega k m}(R,R)\\
&=
-2\pi\int_0^{\ell^2}{ds\over s}\,e^{s\varpi^2-{R^2\over 2 s}}\,I_m\big({R^2\over 2 s}\big).
\end{split}\ee
This function is real and negative.
If the radius of the rotating cylinder $R\gg\ell$, then we obtain
\be\label{gapprox}
g\approx -{2\pi\over\varpi R}\erfi(\varpi\ell).
\ee
In general, $g$ is the function of the parameters $m,\varpi,R,\ell$, but in the limit when $R\gg\ell$, it does not depend on the angular momentum $m$.

In the other approximation, when $\varpi \ell\ll1$ but  $R$ arbitrary,  the integral \eq{DeltaG6} can be evaluated as follows
\be\begin{split}
g\approx &-{2\pi\over m}\Big[1- {\xi^m \,{}_\ins{2}F_\ins{2}\big(m,m+{1\over 2};m+1,2m+1;-\xi \big)\over 2^{2m} m !} \Big]\\
&-{\pi\varpi^2R^2\over m-1}\Big[{1\over m(m+1)}  \\
&- {2\xi^{m-1} \,{}_\ins{2}F_\ins{2}\big(m-1,m+{1\over 2};m,2m+1;-\xi \big)\over 2^{2m} m !} \Big],
\end{split}\ee
where
\be
\xi={R^2\over\ell^2}.
\ee

Similarly to the local case we have
\be
\psi=\psi_\ins{in}+\psi_\ins{out}.
\ee
For $\omega>0$ we have
\be\begin{split}
\psi_\ins{in}&=C H^{(2)}_{m}(\varpi\rho),
\\
\psi_\ins{out}&=C H^{(1)}_{m}(\varpi\rho)-2C{\Lambda\,G_0^\ins{GF}(\rho,R)\over 1+\Lambda G_0^\ins{GF}(R,R)}J_{m},
\end{split}\ee
where
\be
G_0^\ins{GF}(R,R)= i {\pi\over 2}\,J_{m} H^{(1)}_{m}+g.
\ee
The asymptotic of $G_0^\ins{GF}(\rho,R)$ at $\rho\to\infty$  reads
\be
G_0^\ins{GF}(\rho,R)\simeq i {\pi\over 2}\,J_{m} H^{(1)}_{m}(\varpi\rho).
\ee
If the radius of the cylinder is much larger that the scale of nonlocality $R\gg\ell$, then the function $g$ can be approximated by \eq{gapprox}. For small frequencies $\omega\ll \ell^{-1}$ it also leads to a simple asymptotic
\be
g\to -4\sqrt{\pi}{\ell\over R}\Big[1+{1\over 3}\varpi^2\ell^2 +O(\varpi^4\ell^4)\Big].
\ee
Eventually we obtain the asymptotic of $\psi_\ins{out}$ at large $\rho$
\be
\psi_\ins{out}\simeq C H^{(1)}_{m}(\varpi\rho)\Big[1-{ i {\pi\over 2} 2\Lambda J_m^2
\over
1+i{\pi\over 2}\Lambda J_m H^{(1)}_{m}+\Lambda  g
}\Big].
\ee
The complex relative amplitude $A$ is defined as the ratio of amplitudes of the incoming and outgoing waves evaluated at infinity. Thus we obtain
\be
A={1-i{\pi\over 2}\Lambda J_m H^{(2)}_{m}+\Lambda  g
\over
1+i{\pi\over 2}\Lambda J_m H^{(1)}_{m}+\Lambda g}.
\ee
For $\omega<0$, similarly, one gets
\be\begin{split}
\psi_\ins{in}&=C H^{(1)}_{m}(\varpi\rho),
\\
\psi_\ins{out}&=C H^{(2)}_{m}(\varpi\rho)-2C{\Lambda\,G_0^\ins{GF}(\rho,R)\over 1+\Lambda G_0^\ins{GF}(R,R)}J_{m}.
\end{split}\ee
Here
\be
G_0^\ins{GF}(R,R)= - i {\pi\over 2}\,J_{m} H^{(2)}_{m}+g.
\ee
At large $\rho$ we have an asymptotic
\be
G_0^\ins{GF}(\rho,R)\simeq -i {\pi\over 2}\,J_{m} H^{(2)}_{m}(\varpi\rho),
\ee
and, therefore,
\be
\psi_\ins{out}\simeq C H^{(2)}_{m}(\varpi\rho)\Big[1+{ i {\pi\over 2} 2\Lambda J_m^2
\over
1-i{\pi\over 2}\Lambda J_m H^{(2)}_{m}+\Lambda g
}\Big].
\ee
The complex relative amplitude $\widetilde{A}$ is defined as the ratio of amplitudes of the incoming and outgoing waves evaluated at infinity. Thus one gets
\be
\widetilde{A}={1+i{\pi\over 2}\Lambda J_m H^{(1)}_{m}+\Lambda  g
\over
1-i{\pi\over 2}\Lambda J_m H^{(2)}_{m}+\Lambda g}.
\ee
It satisfies the usual relation
\be
\widetilde{A}=A^{-1}.
\ee
Taking into account that $\Lambda=\beta+i\gamma$ one can compute the amplification factor $Z=|A|^2-1$ for positive frequencies
\be\begin{split}\label{Z}
Z&={ 8\gamma \over \pi\Big[
\big(\gamma J_m+\beta Y_m-{2(1+\beta g)\over\pi  J_m}\big)^2+\big(\beta J_m-\gamma Y_m+{2\gamma g\over\pi  J_m}\big)^2\Big]}.
\end{split}\ee

Let us fix the parameters $\ell,R,\varpi$ and find the value of $\gamma$ at which the amplification is maximal. It is easy to find that is happens when
\be
\gamma_\ins{max}={N\over M},
\ee
where
\be\begin{split}
&L=\pi J_m Y_m-2g,\\
&M=\sqrt{\pi^2 (J_m)^4+L^2},\\
&N=\sqrt{(2-\beta L)^2+\beta^2 \pi^2 (J_m)^4}.
\end{split}\ee
Both $M$ and $N$ are positive functions.
The maximal value of $Z$ is then given by
\be
Z_\ins{max}={4\pi (J_m)^2\over MN-2\pi (J_m)^2}.
\ee
The value of $\beta$ which maximizes $Z$ is
\be
\beta_\ins{max}={2 L\over M^2}.
\ee
Remarkably it does not depend on the value of $\gamma$.

In this paper we restrict our consideration with only positive value of the parameter $\beta$, which corresponds to a semitransparent potential barrier, rather than well. In the local theory $g=0$ and $\beta_\ins{max}$ is always negative for $m>0$, $0\le R\le\Omega^{-1}$ and $0\le \varpi R\le m$. The last restriction on $\varpi$ comes from requirement of positivity of $\gamma$, what is equivalent to considering only amplified modes. In fact in the local theory the amplification factor monotonically decreases with growth of the strength $\beta$ of $\delta$-potential. As a result, the maximum amplification occurs when $\beta=0$.
On the other hand, in the non-local theory $\beta_\ins{max}$ may be either negative or positive. In the latter case the maximum of the amplification factor is located not at $\beta=0$ but at $\beta_\ins{max}>0$.

The amplification factor $Z$ diverges when
\be
\beta=\beta_\ins{max}
\ee
and
\be
\gamma=\gamma_\ins{max}(\beta_\ins{max})={2\pi (J_m)^2\over M^2}.
\ee
These values correspond exactly to the complex condition
\be\label{sing}
1+\Lambda G_0^\ins{GF}(R,R)=0.
\ee
This expression naturally enters the Lippmann–Schwinger equation \eq{psiEQ} and is assumed not to be vanishing in the derivation of its solution \eq{Lipp}.\footnote{In the case of a vanishing of damping factor $\alpha=0=\gamma$, the condition \eq{sing} is closely related to the condition of existence of a bound state in the $\delta$-potential and corresponding quasinormal modes \cite{Boos:2018kir}.}


\section{Properties of superradiance amplification}\label{section6}


First of all let us consider superradiance in the local theory, i.e., when $\ell=0$ and hence $g=0$. Let us introduce dimensionless frequency of rotation and mode momentum
\be
 P=\Omega R\hh p=\sqrt{\omega^2-k^2}\, R=\varpi R,
\ee
and substitute
\be
\gamma=\alpha {m P- p\over \sqrt{1-P^2}}
\ee
to the \eq{Z}. The mode is amplified when $\gamma>0$ and, thus, when $p$ is in the interval
\be
0 < p < mP.
\ee

\begin{figure*}[!htb]%
    \centering
    \subfloat{{\includegraphics[width=0.47\textwidth]{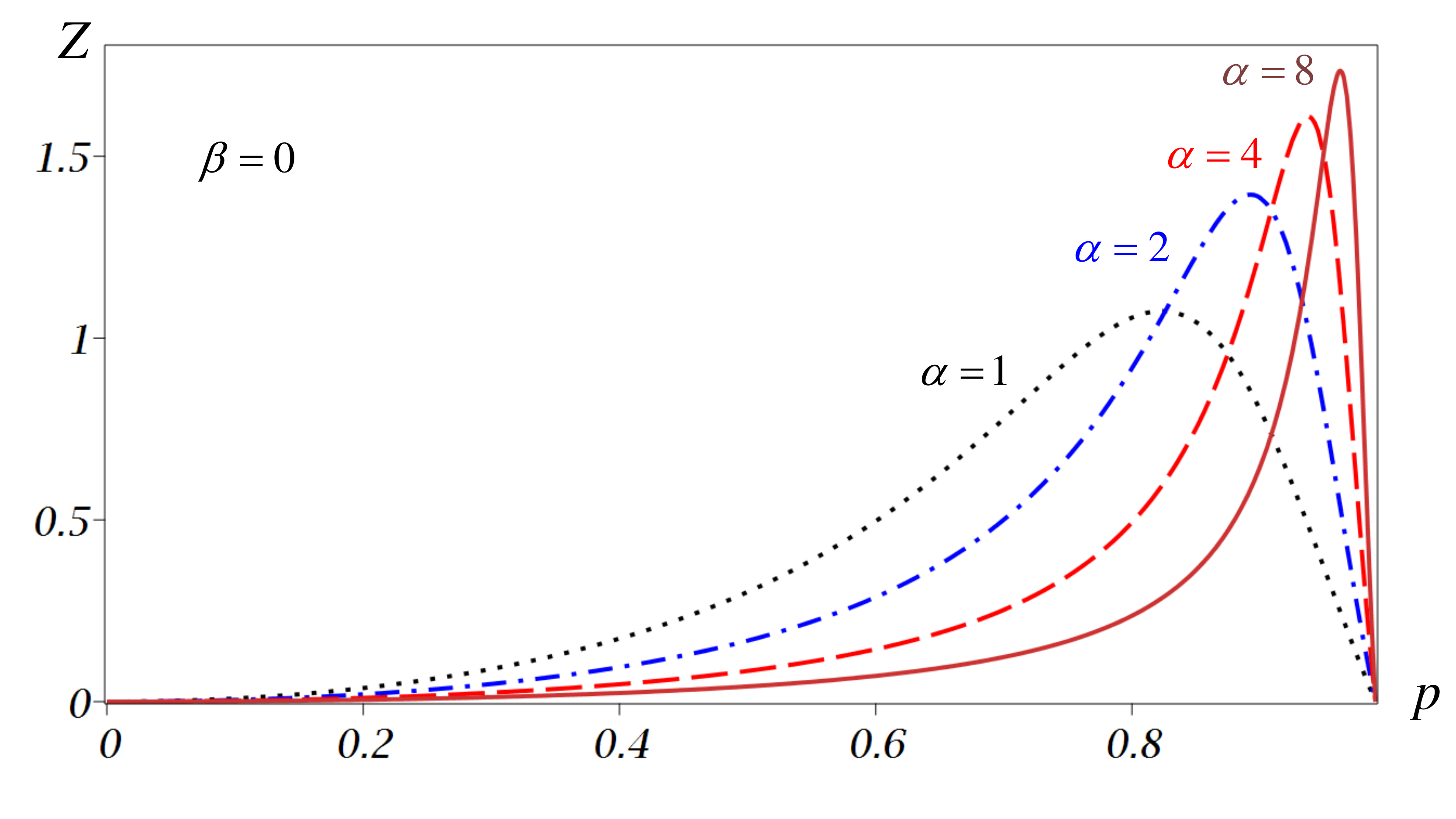} }}%
    \qquad
    \subfloat{{\includegraphics[width=0.47\textwidth]{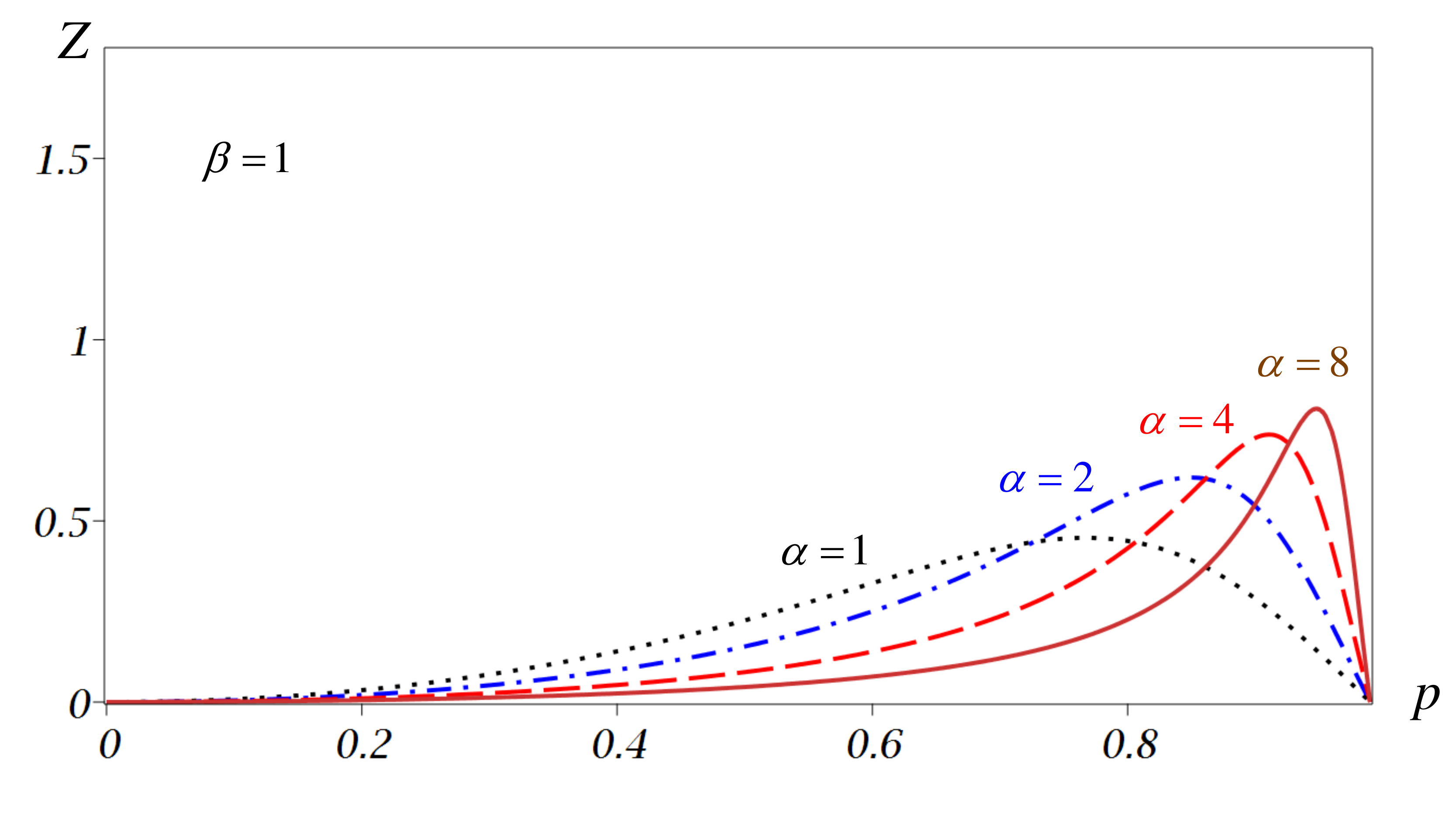} }}
    \caption{The amplification factor $Z(p)$ in the local scalar theory for $\Omega=0.99 \,R^{-1}$, $m=1$, $\beta=0$ (on the left) and $\beta=1 $(on the right). }
    \label{fig1}
\end{figure*}

\begin{figure*}[!htb]%
    \centering
    \subfloat{{\includegraphics[width=0.47\textwidth]{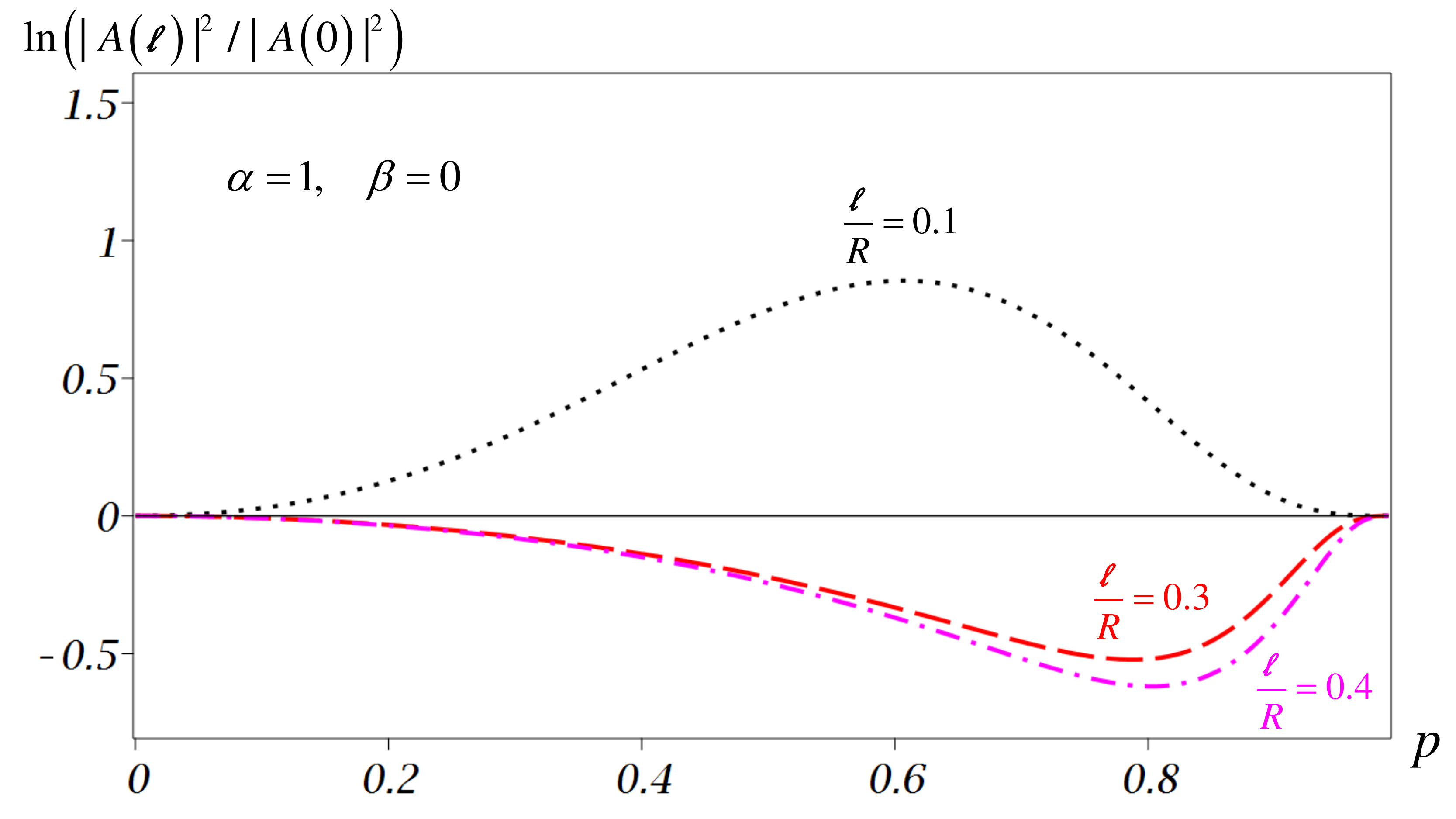} }}%
    \qquad
    \subfloat{{\includegraphics[width=0.47\textwidth]{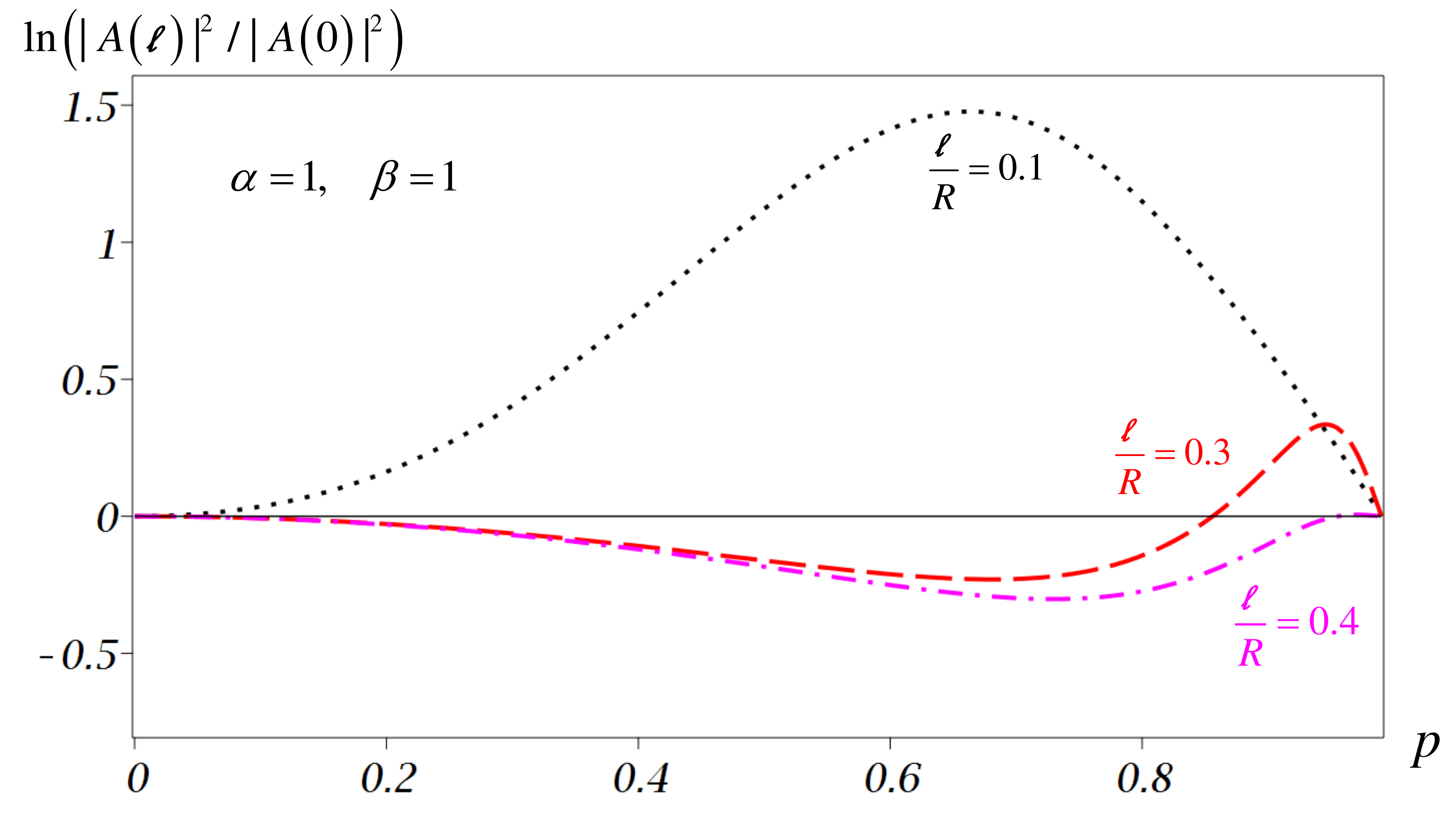} }}
    \caption{Enhancement $\ln[(1+Z)/(1+Z|_{\ell=0})]=\ln[|A|^2/|A|^2_{\ell=0}]$ in the non-local scalar theory for $\Omega=0.99\,R^{-1}$, $m=1$, $\beta=0$ (on the left) and $\beta=1 $(on the right) and a few values of non-locality scale $\ell$. Local theory would corresponds to $\ell=0$. }
    \label{fig2}
\end{figure*}

\begin{figure*}[!htb]%
    \centering
    \subfloat{{\includegraphics[width=0.47\textwidth]{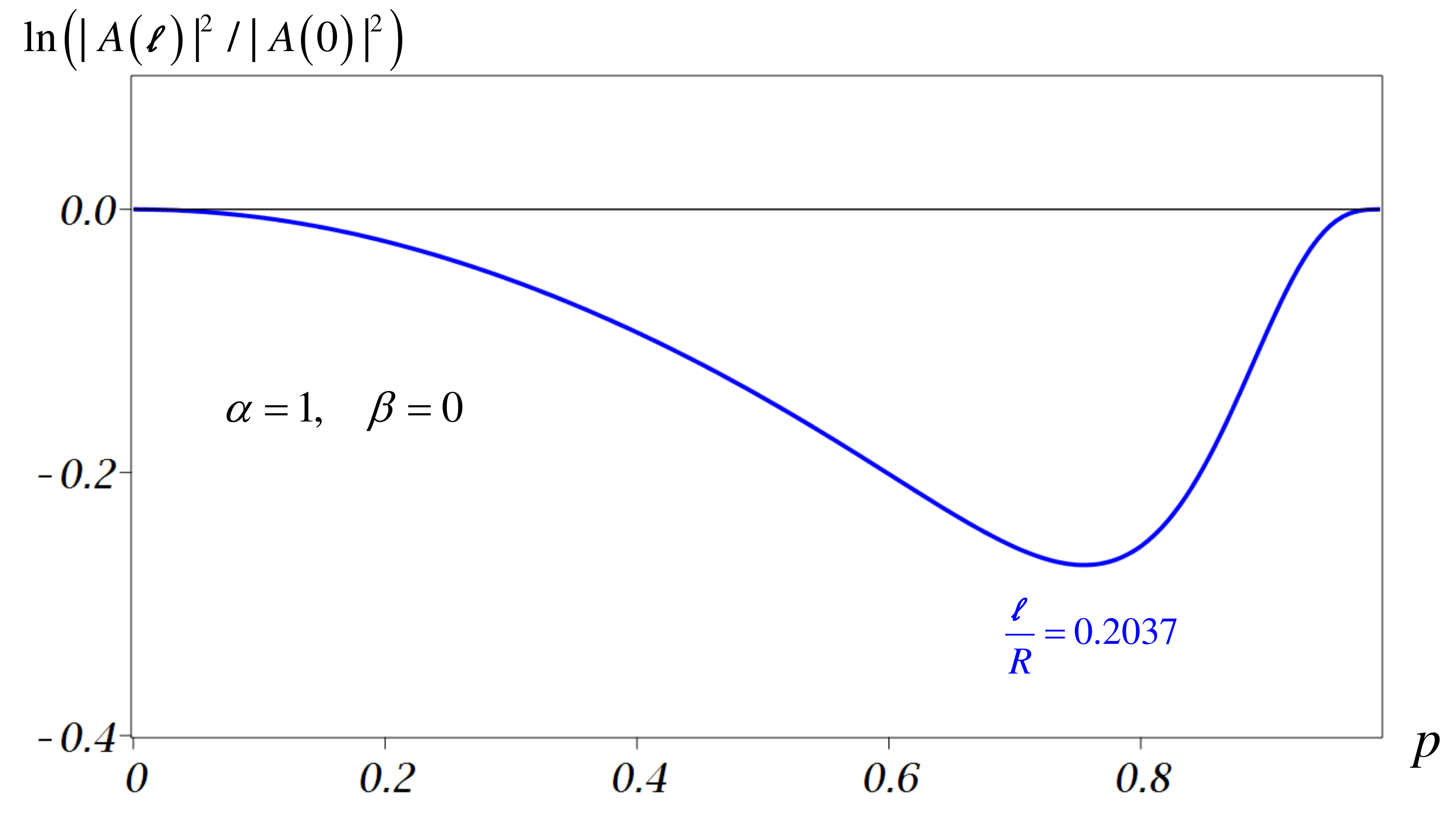} }}%
    \qquad
    \subfloat{{\includegraphics[width=0.47\textwidth]{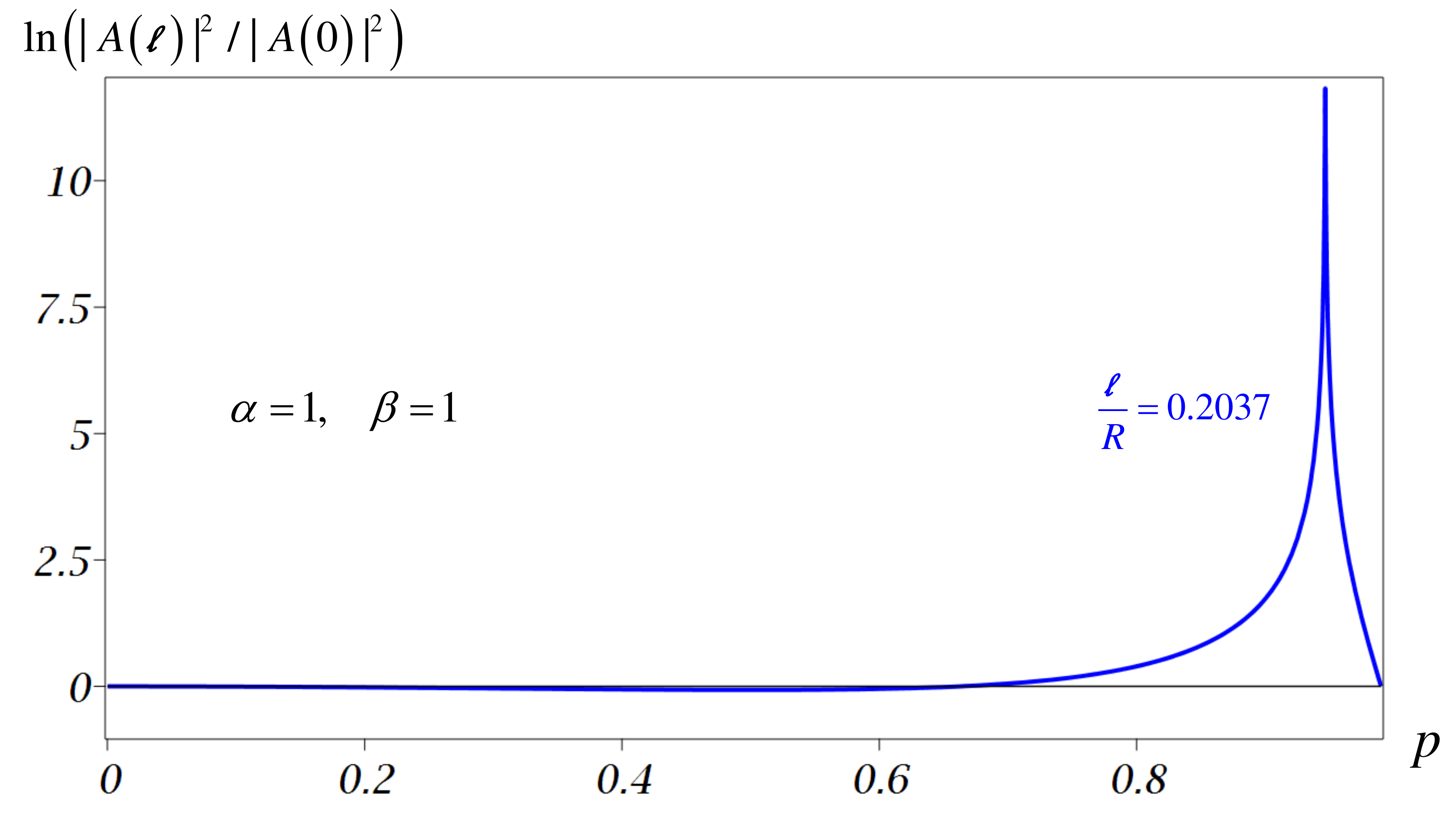} }}
    \caption{Enhancement $\ln[(1+Z)/(1+Z|_{\ell=0})]=\ln[|A|^2/|A|^2_{\ell=0}]$ in the non-local scalar theory for $\Omega=0.99\,R^{-1}$, $m=1$, $\beta=0$ (on the left) and $\beta=1 $(on the right). One can see that at $\ell/R \approx 0.2037$ and $\beta=1$ there is a huge amplification  of the superradiance, about $\sim 1.4\cdot 10^5$ times stronger, as compared to the local theory. This happens at $p\approx 0.9466$. The maximum value of amplification and the corresponding frequency are quite sensitive to the values of $\alpha$, $\beta$, and $\ell/R$.}
    \label{fig3}
\end{figure*}

When $p=mP$ the factor $\gamma$ vanishes and, therefore, $Z$ vanishes as well. When the angular velocity of the cylinder reaches its maximum $P=1$ and the factor $\gamma$ diverges. In this limit amplification factor $Z$ vanishes both in local and non-local theories. The strongest amplification is achieved typically, but not always, for the high angular velocities of the cylinder. Now consider dependence of  $Z$ on the momentum parameter $p$. At small $p$ the Bessel function $J_m(p)\approx p^m2^{-m}/m!$. Thus for $m\ge1$ amplification also vanishes at $p=0$.

\begin{figure*}[!htb]%
    {\includegraphics[width=0.47\textwidth]{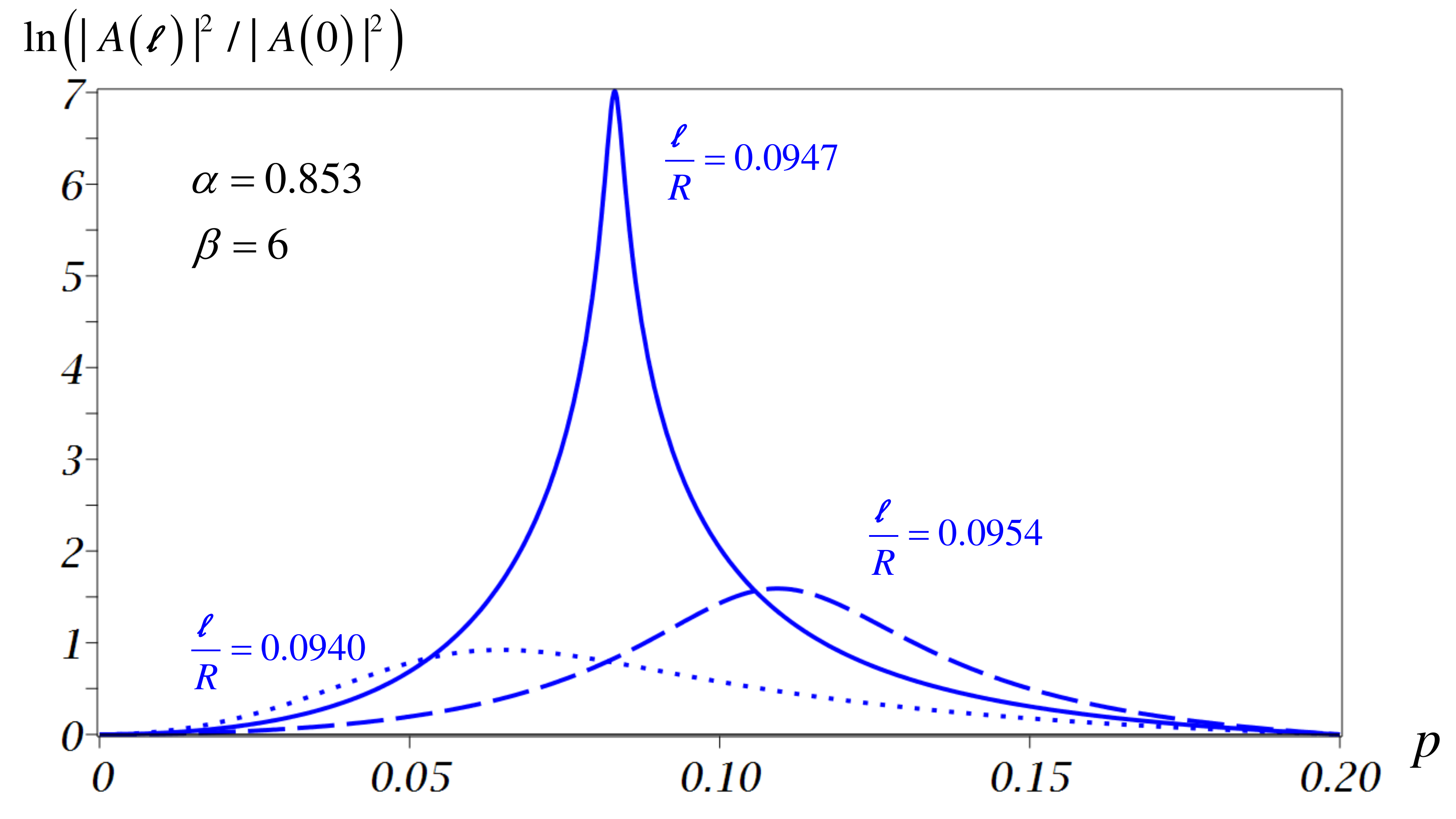}}%
    \caption{Enhancement $\ln[(1+Z)/(1+Z|_{\ell=0})]=\ln[|A|^2/|A|^2_{\ell=0}]$ in the non-local scalar theory for moderate values of $\Omega=0.2\,R^{-1}$, $m=1$, $\alpha=0.853$ and $\beta=6$. The strongest resonance amplification is at $\ell/R\approx 0.0947$ (solid live). One can see, that even a tiny variation of the parameter $\ell/R$ significantly changes both the amplitude of the amplification and the characteristic frequency (see dotted and dashed lines). }
    \label{fig4}
\end{figure*}

Let us discuss now how these results, obtained for the local theory, are modified when we include the non-locality effects. To simplify the presentation, in what follows we put  $m=1$ and $k=0$.\footnote{The theory without absorption is invariant under the boosts along $z$ axis. In this case one can put $k=0$ without loss of generality. However, in the case of a non-vanishing absorption this symmetry is broken and there appears dependence on $k$ via the combination $\sqrt{\omega^2-k^2}$.}.  In Fig.~\ref{fig1} we depicted typical frequency dependence of the amplification factor $Z(p)$. In order to demonstrate how nonlocality affects the superradiance we present plots with a few ratios of the radius $R$ of a rotating cylinder to the non-locality scale $\ell$. In Fig.~\ref{fig2} one can see that for $\ell/R=(0.1,0.3,0.4)$ the effect of nonlocality is quite modest, the superradiance is of the same order of magnitude. But when $\ell/R\approx 0.2$, the nonlocality leads to a huge  enhancement of superradiance (see Fig.~\ref{fig3}). For example when  $\ell/R \approx 0.2037$ the non-local superradiance about $10^5$ times stronger than in the local theory. There is a range of parameters, where a resonant non-local amplification of superradiance occurs. Usually it happens in a quite narrow corridor in the parameter space. A example of a non-locally enhanced superradiance for quite low rotations $\Omega=0.2 R^{-1}$ and at moderate frequencies $\omega\sim 0.1 R^{-1}$ is depicted in Fig.~\ref{fig4}.

\section{Discussion}\label{section7}

Let us summarize the results. We considered scattering of a ghost-free scalar massless field by a rotating (with the angular velocity $\Omega$) cylinder, and demonstrated that in the presence of absorption of the wave by the cylinder, its amplitude can be amplified. It happens for the same superradiance condition as for the local case $0<\omega<m \Omega$. However, the dependence of the amplification coefficient on the frequency of the wave, as well as the parameters characterizing the cylinder (its height, $\beta$, and its absorption factor, $\alpha$), might considerably differ from the local case. We demonstrated that for a chosen model of thin and empty inside cylinder the scattering problem in both (local and non-local) cases is exactly solvable. The required solution was found by using the Lippmann-Schwinger equation. The solution contains cylindrical harmonics of the free retarded Green function of the corresponding problem in the absence of the potential. The modification of the amplification coefficient in the ghost-free case is related with a non-local contribution into the retarded Green function. We demonstrated the dependence of this coefficient on the parameter of the non-locality $\ell$. In particular, we found that the superradiance effect can be greatly amplified for special relation between the frequency of the radiation, parameters of the potential and of the non-locality. A similar effect was observed earlier in the scattering of a ghost-free wave on the delta-like potential barrier \cite{Boos:2018kir}. We studied superradiance in the simplest case of $GF_1$ theory. It has an advantage that all computations can be performed analytically and the result can presented in an explicit form. Disadvantage of $GF_1$ theory is that it suffers instabilities at very high frequencies $\omega\gg\ell^{-1}$ (see \cite{Frolov:2016xhq}). However, for consideration of superradiance this drawback is not of our concern, because superradiance is important at low frequencies.  $GF_N$ theories with even $N$ do not suffer with this instability and at the same time one can expect that superradiance qualitatively behaves in a very similar manner.

Zel'dovich \cite{zeldovich:1971} used the analogy between a scattering of the waves on a rotating absorbing cylinder and the scattering of similar waves on a rotating black hole in order to predict the effect of the black hole induced superradiance. An interesting question is: Is this analogy also valid for the ghost-free field. This point requires explanations. In the case of a rotating absorbing cylinder, which we considered in this paper, the potential describing the cylinder modifies the free propagation of the field. In other words, this is a so-called off-shell problem and the corresponding retarded Green function, which is used to describe this effect, is a solution of the inhomogeneous equation with the delta-function in its right-hand side. Such Green functions are different for the local and non-local cases (see e.g. \cite{Buoninfante:2018xiw}). In the case of a rotating black hole, the background metric enters the form-factor of the non-locality, so that, at least at first glance, for the description of the wave scattering by the black hole it is sufficient to use only on-shell quantities, which are identical for the local and ghost-free non-local cases. This would imply, that the superradiance is also valid for the ghost-free field, but the effects of the non-locality do not manifest themselves in this case.
Anyway, it would be very interesting to perform accurate calculations of the super-radiant scattering of the ghost-free field on a rotating black hole and to check whether the Zel'dovich's analogy is still valid and complete.

\section*{Acknowledgments}

The authors thank the Natural Sciences and Engineering Research Council of Canada and the Killam Trust for their financial support.

\appendix

\section{Absorption on $\delta$-potential}\label{absorption}

Let us consider a complex massless scalar field $\varphi$ in a curved spacetime interacting with an absorbing medium concentrated on a surface $\Sigma$. It obeys the equation
\be \n{a.1}
\Box \varphi -V\varphi=0 ,
\ee
where
\be\n{a.2}
V=\delta(\Sigma)[b+a u^{\mu}\nabla_{\mu}]\, .
\ee
$\Sigma$ is a timelike surface and $u^{\mu}$ is a unit future-directed timelike vector tangent to $\Sigma$. The real coefficients $b$ and $a$ characterize the ``height" of the $\delta$-potential and its ``absorption capacity", respectively. Let us denote
\be\n{a.3}
j_{\mu}=-i[\varphi^* \nabla_{\mu} \varphi -\varphi_{\mu} \nabla \varphi^*]\, .
\ee
Equation (\ref{a.1}) and its conjugated imply
\be\n{a.4}
\nabla^{\mu}j_{\mu}-a u^{\mu}j_{\mu} \delta(\Sigma)=0\, .
\ee

Let $\sigma_1$ and $\sigma_2$ be two spacelike surfaces. Let us assume that current $j_{\mu}$ vanishes fast enough at spatial infinity, then integrating over the 4-volume $V$ between $\sigma_1$ and $\sigma_2$ and using the Stokes theorem one obtains
\be\n{a.5}
Q[\sigma_2]-Q[\sigma_1]=-a \int dv u^{\mu} j_{\mu} \delta(\Sigma)\, .
\ee
Here
\be\n{a.6}
Q[\sigma]=-\int_{\sigma} n^{\mu} j_{\mu} \sqrt{h} d^3y\, .
\ee
$\bf{h}$ is the induced metric on $\sigma$ and $n^{\mu}$ is a future-directed unit normal to $\sigma$ vector.

The equation (\ref{a.5}) has a simple meaning. Namely, the total charge $Q$, carried by the field $\varphi$ changes with time as a result of its absorption by the potential $V$. For a real quantized scalar field in the flat spacetime one can relate $\varphi$ and $\varphi^*$ with its positive and negative frequency parts. In this case a relation similar to (\ref{a.5}) would describe the absorption of the scalar quanta by the potential $V$.

Let us adapt the obtained relations to a simple case of the 4-dimensional flat spacetime. Let its Cartesian coordinates be $(T,X,Y,Z)$ and the surface $\Sigma$ is a 3-plane $Z=0$. One has
\be\n{a.7}
u^{\mu}\partial_{\mu}=\partial_T\, .
\ee
We choose $\sigma$ as a surface $T=$const, so that $n^{\mu}\partial_{\mu}=\partial_T$ and
\be\n{a.8}
Q[T]=\int dX dY dZ \, j^T ,
\ee
and one obtains
\be\n{a.9}
{dQ\over dT}=-a\int j^T\big|_\ins{Z} \,dX dY\, .
\ee
This relation shows that the rate of the change of ``number of particles" is proportional to the ``particle density" in the vicinity of the barrier, and the coefficient $a$ characterizes how fast this process is.



\begin{thebibliography}{36}%
\makeatletter
\providecommand \@ifxundefined [1]{%
 \@ifx{#1\undefined}
}%
\providecommand \@ifnum [1]{%
 \ifnum #1\expandafter \@firstoftwo
 \else \expandafter \@secondoftwo
 \fi
}%
\providecommand \@ifx [1]{%
 \ifx #1\expandafter \@firstoftwo
 \else \expandafter \@secondoftwo
 \fi
}%
\providecommand \natexlab [1]{#1}%
\providecommand \enquote  [1]{``#1''}%
\providecommand \bibnamefont  [1]{#1}%
\providecommand \bibfnamefont [1]{#1}%
\providecommand \citenamefont [1]{#1}%
\providecommand \href@noop [0]{\@secondoftwo}%
\providecommand \href [0]{\begingroup \@sanitize@url \@href}%
\providecommand \@href[1]{\@@startlink{#1}\@@href}%
\providecommand \@@href[1]{\endgroup#1\@@endlink}%
\providecommand \@sanitize@url [0]{\catcode `\\12\catcode `\$12\catcode
  `\&12\catcode `\#12\catcode `\^12\catcode `\_12\catcode `\%12\relax}%
\providecommand \@@startlink[1]{}%
\providecommand \@@endlink[0]{}%
\providecommand \url  [0]{\begingroup\@sanitize@url \@url }%
\providecommand \@url [1]{\endgroup\@href {#1}{\urlprefix }}%
\providecommand \urlprefix  [0]{URL }%
\providecommand \Eprint [0]{\href }%
\providecommand \doibase [0]{http://dx.doi.org/}%
\providecommand \selectlanguage [0]{\@gobble}%
\providecommand \bibinfo  [0]{\@secondoftwo}%
\providecommand \bibfield  [0]{\@secondoftwo}%
\providecommand \translation [1]{[#1]}%
\providecommand \BibitemOpen [0]{}%
\providecommand \bibitemStop [0]{}%
\providecommand \bibitemNoStop [0]{.\EOS\space}%
\providecommand \EOS [0]{\spacefactor3000\relax}%
\providecommand \BibitemShut  [1]{\csname bibitem#1\endcsname}%
\let\auto@bib@innerbib\@empty
\bibitem [{\citenamefont {Zel'dovich}(1971)}]{zeldovich:1971}%
  \BibitemOpen
  \bibfield  {author} {\bibinfo {author} {\bibfnamefont {Y.~B.}\ \bibnamefont
  {Zel'dovich}},\ }\bibfield  {title} {\emph {\enquote {\bibinfo {title}
  {Generation of waves by a rotating body},}\ }}\href@noop {} {\bibfield
  {journal} {\bibinfo  {journal} {Sov. Phys. JETP Letters}\ }\textbf {\bibinfo
  {volume} {14}},\ \bibinfo {pages} {180} (\bibinfo {year} {1971})}\BibitemShut
  {NoStop}%
\bibitem [{\citenamefont {Zel'dovich}(1972)}]{zeldovich:1972}%
  \BibitemOpen
  \bibfield  {author} {\bibinfo {author} {\bibfnamefont {Y.~B.}\ \bibnamefont
  {Zel'dovich}},\ }\bibfield  {title} {\emph {\enquote {\bibinfo {title}
  {Amplification of cylindrical electromagnetic waves reflected from a rotating
  body},}\ }}\href@noop {} {\bibfield  {journal} {\bibinfo  {journal} {Sov.
  Phys. JETP}\ }\textbf {\bibinfo {volume} {35}},\ \bibinfo {pages} {1085}
  (\bibinfo {year} {1972})}\BibitemShut {NoStop}%
\bibitem [{\citenamefont {Bolotovskii}\ and\ \citenamefont
  {Stolyarov}(1975)}]{Bolotovskii:1975}%
  \BibitemOpen
  \bibfield  {author} {\bibinfo {author} {\bibfnamefont {B.~M.}\ \bibnamefont
  {Bolotovskii}}\ and\ \bibinfo {author} {\bibfnamefont {S.~N.}\ \bibnamefont
  {Stolyarov}},\ }\bibfield  {title} {\emph {\enquote {\bibinfo {title}
  {Current status of the electrodynamics of moving media (infinite media)},}\
  }}\href {\doibase 10.1070/PU1975v017n06ABEH004403} {\bibfield  {journal}
  {\bibinfo  {journal} {Phys. Usp.}\ }\textbf {\bibinfo {volume} {18}},\
  \bibinfo {pages} {875} (\bibinfo {year} {1975})}\BibitemShut {NoStop}%
\bibitem [{\citenamefont {Starobinsky}(1973)}]{Starobinsky:1973aij}%
  \BibitemOpen
  \bibfield  {author} {\bibinfo {author} {\bibfnamefont {A.~A.}\ \bibnamefont
  {Starobinsky}},\ }\bibfield  {title} {\emph {\enquote {\bibinfo {title}
  {{Amplification of waves reflected from a rotating "black hole".}}}\
  }}\href@noop {} {\bibfield  {journal} {\bibinfo  {journal} {Sov. Phys. JETP}\
  }\textbf {\bibinfo {volume} {37}},\ \bibinfo {pages} {28} (\bibinfo {year}
  {1973})},\ \bibinfo {note} {[Zh. Eksp. Teor. Fiz.64,48(1973)]}\BibitemShut
  {NoStop}%
\bibitem [{\citenamefont {Starobinskii}\ and\ \citenamefont
  {Churilov}(1974)}]{Starobinsky:1974nkd}%
  \BibitemOpen
  \bibfield  {author} {\bibinfo {author} {\bibfnamefont {A.~A.}\ \bibnamefont
  {Starobinskii}}\ and\ \bibinfo {author} {\bibfnamefont {S.~M.}\ \bibnamefont
  {Churilov}},\ }\bibfield  {title} {\emph {\enquote {\bibinfo {title}
  {{Amplification of electromagnetic and gravitational waves scattered by a
  rotating "black hole"}},}\ }}\href@noop {} {\bibfield  {journal} {\bibinfo
  {journal} {Sov. Phys. JETP}\ }\textbf {\bibinfo {volume} {65}},\ \bibinfo
  {pages} {1} (\bibinfo {year} {1974})}\BibitemShut {NoStop}%
\bibitem [{\citenamefont {Breuer}\ \emph {et~al.}(1973)\citenamefont {Breuer},
  \citenamefont {Chrzanowksi}, \citenamefont {Hughes},\ and\ \citenamefont
  {Misner}}]{Breuer:1973uc}%
  \BibitemOpen
  \bibfield  {author} {\bibinfo {author} {\bibfnamefont {R.~A.}\ \bibnamefont
  {Breuer}}, \bibinfo {author} {\bibfnamefont {P.~L.}\ \bibnamefont
  {Chrzanowksi}}, \bibinfo {author} {\bibfnamefont {H.~G.}\ \bibnamefont
  {Hughes}}, \ and\ \bibinfo {author} {\bibfnamefont {C.~W.}\ \bibnamefont
  {Misner}},\ }\bibfield  {title} {\emph {\enquote {\bibinfo {title} {{Geodesic
  synchrotron radiation}},}\ }}\href {\doibase 10.1103/PhysRevD.8.4309}
  {\bibfield  {journal} {\bibinfo  {journal} {Phys. Rev.}\ }\textbf {\bibinfo
  {volume} {D8}},\ \bibinfo {pages} {4309} (\bibinfo {year}
  {1973})}\BibitemShut {NoStop}%
\bibitem [{\citenamefont {Unruh}(1974)}]{Unruh:1974bw}%
  \BibitemOpen
  \bibfield  {author} {\bibinfo {author} {\bibfnamefont {W.~G.}\ \bibnamefont
  {Unruh}},\ }\bibfield  {title} {\emph {\enquote {\bibinfo {title} {{Second
  quantization in the Kerr metric}},}\ }}\href {\doibase
  10.1103/PhysRevD.10.3194} {\bibfield  {journal} {\bibinfo  {journal} {Phys.
  Rev.}\ }\textbf {\bibinfo {volume} {D10}},\ \bibinfo {pages} {3194} (\bibinfo
  {year} {1974})}\BibitemShut {NoStop}%
\bibitem [{\citenamefont {Ginzburg}(1993)}]{Ginzburg:1993}%
  \BibitemOpen
  \bibfield  {author} {\bibinfo {author} {\bibfnamefont {V.~L.}\ \bibnamefont
  {Ginzburg}},\ }in\ \href@noop {} {\emph {\bibinfo {booktitle} {{Progress in
  optics}}}},\ Vol.~\bibinfo {volume} {32},\ \bibinfo {editor} {edited by\
  \bibinfo {editor} {\bibfnamefont {E.}~\bibnamefont {Wolf}}}\ (\bibinfo
  {publisher} {{Elsevier}},\ \bibinfo {address} {{Amsterdam}},\ \bibinfo {year}
  {1993})\ pp.\ \bibinfo {pages} {267--312}\BibitemShut {NoStop}%
\bibitem [{\citenamefont {Bekenstein}\ and\ \citenamefont
  {Schiffer}(1998)}]{Bekenstein:1998nt}%
  \BibitemOpen
  \bibfield  {author} {\bibinfo {author} {\bibfnamefont {J.~D.}\ \bibnamefont
  {Bekenstein}}\ and\ \bibinfo {author} {\bibfnamefont {M.}~\bibnamefont
  {Schiffer}},\ }\bibfield  {title} {\emph {\enquote {\bibinfo {title} {{The
  Many faces of superradiance}},}\ }}\href {\doibase
  10.1103/PhysRevD.58.064014} {\bibfield  {journal} {\bibinfo  {journal} {Phys.
  Rev.}\ }\textbf {\bibinfo {volume} {D58}},\ \bibinfo {pages} {064014}
  (\bibinfo {year} {1998})},\ \Eprint {http://arxiv.org/abs/gr-qc/9803033}
  {arXiv:gr-qc/9803033 [gr-qc]} \BibitemShut {NoStop}%
\bibitem [{\citenamefont {Frolov}\ and\ \citenamefont
  {Ginzburg}(1986)}]{Frolov:1986}%
  \BibitemOpen
  \bibfield  {author} {\bibinfo {author} {\bibfnamefont {V.~P.}\ \bibnamefont
  {Frolov}}\ and\ \bibinfo {author} {\bibfnamefont {V.~L.}\ \bibnamefont
  {Ginzburg}},\ }\bibfield  {title} {\emph {\enquote {\bibinfo {title}
  {Excitation and radiation of an accelerated detector and anomalous doppler
  effect},}\ }}\href {http://stacks.iop.org/0038-5670/30/i=12/a=A04} {\bibfield
   {journal} {\bibinfo  {journal} {Phys. Lett. A}\ }\textbf {\bibinfo {volume}
  {116}},\ \bibinfo {pages} {423} (\bibinfo {year} {1986})}\BibitemShut
  {NoStop}%
\bibitem [{\citenamefont {Ginzburg}\ and\ \citenamefont
  {Frolov}(1987)}]{Ginzburg:1987}%
  \BibitemOpen
  \bibfield  {author} {\bibinfo {author} {\bibfnamefont {V.~L.}\ \bibnamefont
  {Ginzburg}}\ and\ \bibinfo {author} {\bibfnamefont {V.~P.}\ \bibnamefont
  {Frolov}},\ }\bibfield  {title} {\emph {\enquote {\bibinfo {title} {Vacuum in
  a homogeneous gravitational field and excitation of a uniformly accelerated
  detector},}\ }}\href {http://stacks.iop.org/0038-5670/30/i=12/a=A04}
  {\bibfield  {journal} {\bibinfo  {journal} {Soviet Physics Uspekhi}\ }\textbf
  {\bibinfo {volume} {30}},\ \bibinfo {pages} {1073} (\bibinfo {year}
  {1987})}\BibitemShut {NoStop}%
\bibitem [{\citenamefont {Torres}\ \emph {et~al.}(2017)\citenamefont {Torres},
  \citenamefont {Patrick}, \citenamefont {Coutant}, \citenamefont {Richartz},
  \citenamefont {Tedford},\ and\ \citenamefont {Weinfurtner}}]{Torres:2016iee}%
  \BibitemOpen
  \bibfield  {author} {\bibinfo {author} {\bibfnamefont {T.}~\bibnamefont
  {Torres}}, \bibinfo {author} {\bibfnamefont {S.}~\bibnamefont {Patrick}},
  \bibinfo {author} {\bibfnamefont {A.}~\bibnamefont {Coutant}}, \bibinfo
  {author} {\bibfnamefont {M.}~\bibnamefont {Richartz}}, \bibinfo {author}
  {\bibfnamefont {E.~W.}\ \bibnamefont {Tedford}}, \ and\ \bibinfo {author}
  {\bibfnamefont {S.}~\bibnamefont {Weinfurtner}},\ }\bibfield  {title} {\emph
  {\enquote {\bibinfo {title} {{Observation of superradiance in a vortex
  flow}},}\ }}\href {\doibase 10.1038/nphys4151} {\bibfield  {journal}
  {\bibinfo  {journal} {Nature Phys.}\ }\textbf {\bibinfo {volume} {13}},\
  \bibinfo {pages} {833} (\bibinfo {year} {2017})},\ \Eprint
  {http://arxiv.org/abs/1612.06180} {arXiv:1612.06180 [gr-qc]} \BibitemShut
  {NoStop}%
\bibitem [{\citenamefont {Cardoso}\ \emph {et~al.}(2016)\citenamefont
  {Cardoso}, \citenamefont {Coutant}, \citenamefont {Richartz},\ and\
  \citenamefont {Weinfurtner}}]{Cardoso:2016zvz}%
  \BibitemOpen
  \bibfield  {author} {\bibinfo {author} {\bibfnamefont {V.}~\bibnamefont
  {Cardoso}}, \bibinfo {author} {\bibfnamefont {A.}~\bibnamefont {Coutant}},
  \bibinfo {author} {\bibfnamefont {M.}~\bibnamefont {Richartz}}, \ and\
  \bibinfo {author} {\bibfnamefont {S.}~\bibnamefont {Weinfurtner}},\
  }\bibfield  {title} {\emph {\enquote {\bibinfo {title} {{Detecting Rotational
  Superradiance in Fluid Laboratories}},}\ }}\href {\doibase
  10.1103/PhysRevLett.117.271101} {\bibfield  {journal} {\bibinfo  {journal}
  {Phys. Rev. Lett.}\ }\textbf {\bibinfo {volume} {117}},\ \bibinfo {pages}
  {271101} (\bibinfo {year} {2016})},\ \Eprint
  {http://arxiv.org/abs/1607.01378} {arXiv:1607.01378 [gr-qc]} \BibitemShut
  {NoStop}%
\bibitem [{\citenamefont {Richartz}\ \emph {et~al.}(2009)\citenamefont
  {Richartz}, \citenamefont {Weinfurtner}, \citenamefont {Penner},\ and\
  \citenamefont {Unruh}}]{Richartz:2009mi}%
  \BibitemOpen
  \bibfield  {author} {\bibinfo {author} {\bibfnamefont {M.}~\bibnamefont
  {Richartz}}, \bibinfo {author} {\bibfnamefont {S.}~\bibnamefont
  {Weinfurtner}}, \bibinfo {author} {\bibfnamefont {A.~J.}\ \bibnamefont
  {Penner}}, \ and\ \bibinfo {author} {\bibfnamefont {W.~G.}\ \bibnamefont
  {Unruh}},\ }\bibfield  {title} {\emph {\enquote {\bibinfo {title} {{General
  universal superradiant scattering}},}\ }}\href {\doibase
  10.1103/PhysRevD.80.124016} {\bibfield  {journal} {\bibinfo  {journal} {Phys.
  Rev.}\ }\textbf {\bibinfo {volume} {D80}},\ \bibinfo {pages} {124016}
  (\bibinfo {year} {2009})},\ \Eprint {http://arxiv.org/abs/0909.2317}
  {arXiv:0909.2317 [gr-qc]} \BibitemShut {NoStop}%
\bibitem [{\citenamefont {Vicente}\ \emph {et~al.}(2018)\citenamefont
  {Vicente}, \citenamefont {Cardoso},\ and\ \citenamefont
  {Lopes}}]{Vicente:2018mxl}%
  \BibitemOpen
  \bibfield  {author} {\bibinfo {author} {\bibfnamefont {R.}~\bibnamefont
  {Vicente}}, \bibinfo {author} {\bibfnamefont {V.}~\bibnamefont {Cardoso}}, \
  and\ \bibinfo {author} {\bibfnamefont {J.~C.}\ \bibnamefont {Lopes}},\
  }\bibfield  {title} {\emph {\enquote {\bibinfo {title} {{Penrose process,
  superradiance, and ergoregion instabilities}},}\ }}\href {\doibase
  10.1103/PhysRevD.97.084032} {\bibfield  {journal} {\bibinfo  {journal} {Phys.
  Rev.}\ }\textbf {\bibinfo {volume} {D97}},\ \bibinfo {pages} {084032}
  (\bibinfo {year} {2018})},\ \Eprint {http://arxiv.org/abs/1803.08060}
  {arXiv:1803.08060 [gr-qc]} \BibitemShut {NoStop}%
\bibitem [{\citenamefont {Brito}\ \emph {et~al.}(2015)\citenamefont {Brito},
  \citenamefont {Cardoso},\ and\ \citenamefont {Pani}}]{Brito:2015oca}%
  \BibitemOpen
  \bibfield  {author} {\bibinfo {author} {\bibfnamefont {R.}~\bibnamefont
  {Brito}}, \bibinfo {author} {\bibfnamefont {V.}~\bibnamefont {Cardoso}}, \
  and\ \bibinfo {author} {\bibfnamefont {P.}~\bibnamefont {Pani}},\ }\bibfield
  {title} {\emph {\enquote {\bibinfo {title} {{Superradiance}},}\ }}\href
  {\doibase 10.1007/978-3-319-19000-6} {\bibfield  {journal} {\bibinfo
  {journal} {Lect. Notes Phys.}\ }\textbf {\bibinfo {volume} {906}},\ \bibinfo
  {pages} {pp.1} (\bibinfo {year} {2015})},\ \Eprint
  {http://arxiv.org/abs/1501.06570} {arXiv:1501.06570 [gr-qc]} \BibitemShut
  {NoStop}%
\bibitem [{\citenamefont {Frolov}\ and\ \citenamefont
  {Zelnikov}(2016{\natexlab{a}})}]{Frolov:2016xhq}%
  \BibitemOpen
  \bibfield  {author} {\bibinfo {author} {\bibfnamefont {V.~P.}\ \bibnamefont
  {Frolov}}\ and\ \bibinfo {author} {\bibfnamefont {A.}~\bibnamefont
  {Zelnikov}},\ }\bibfield  {title} {\emph {\enquote {\bibinfo {title}
  {{Radiation from an emitter in the ghost free scalar theory}},}\ }}\href
  {\doibase 10.1103/PhysRevD.93.105048} {\bibfield  {journal} {\bibinfo
  {journal} {Phys. Rev.}\ }\textbf {\bibinfo {volume} {D93}},\ \bibinfo {pages}
  {105048} (\bibinfo {year} {2016}{\natexlab{a}})},\ \Eprint
  {http://arxiv.org/abs/1603.00826} {arXiv:1603.00826 [hep-th]} \BibitemShut
  {NoStop}%
\bibitem [{\citenamefont {Buoninfante}\ \emph
  {et~al.}(2018{\natexlab{a}})\citenamefont {Buoninfante}, \citenamefont
  {Lambiase},\ and\ \citenamefont {Mazumdar}}]{Buoninfante:2018mre}%
  \BibitemOpen
  \bibfield  {author} {\bibinfo {author} {\bibfnamefont {L.}~\bibnamefont
  {Buoninfante}}, \bibinfo {author} {\bibfnamefont {G.}~\bibnamefont
  {Lambiase}}, \ and\ \bibinfo {author} {\bibfnamefont {A.}~\bibnamefont
  {Mazumdar}},\ }\bibfield  {title} {\emph {\enquote {\bibinfo {title}
  {{Ghost-free infinite derivative quantum field theory}},}\ }}\href@noop {} {\
   (\bibinfo {year} {2018}{\natexlab{a}})},\ \Eprint
  {http://arxiv.org/abs/1805.03559} {arXiv:1805.03559 [hep-th]} \BibitemShut
  {NoStop}%
\bibitem [{\citenamefont {Buoninfante}\ \emph
  {et~al.}(2018{\natexlab{b}})\citenamefont {Buoninfante}, \citenamefont
  {Cornell}, \citenamefont {Harmsen}, \citenamefont {Koshelev}, \citenamefont
  {Lambiase}, \citenamefont {Marto},\ and\ \citenamefont
  {Mazumdar}}]{Buoninfante:2018xif}%
  \BibitemOpen
  \bibfield  {author} {\bibinfo {author} {\bibfnamefont {L.}~\bibnamefont
  {Buoninfante}}, \bibinfo {author} {\bibfnamefont {A.~S.}\ \bibnamefont
  {Cornell}}, \bibinfo {author} {\bibfnamefont {G.}~\bibnamefont {Harmsen}},
  \bibinfo {author} {\bibfnamefont {A.~S.}\ \bibnamefont {Koshelev}}, \bibinfo
  {author} {\bibfnamefont {G.}~\bibnamefont {Lambiase}}, \bibinfo {author}
  {\bibfnamefont {J.}~\bibnamefont {Marto}}, \ and\ \bibinfo {author}
  {\bibfnamefont {A.}~\bibnamefont {Mazumdar}},\ }\bibfield  {title} {\emph
  {\enquote {\bibinfo {title} {{Non-singular rotating metric in ghost-free
  infinite derivative gravity}},}\ }}\href@noop {} {\  (\bibinfo {year}
  {2018}{\natexlab{b}})},\ \Eprint {http://arxiv.org/abs/1807.08896}
  {arXiv:1807.08896 [gr-qc]} \BibitemShut {NoStop}%
\bibitem [{\citenamefont {Modesto}\ and\ \citenamefont
  {Rachwał}(2017)}]{Modesto:2017sdr}%
  \BibitemOpen
  \bibfield  {author} {\bibinfo {author} {\bibfnamefont {L.}~\bibnamefont
  {Modesto}}\ and\ \bibinfo {author} {\bibfnamefont {L.}~\bibnamefont
  {Rachwał}},\ }\bibfield  {title} {\emph {\enquote {\bibinfo {title}
  {{Nonlocal quantum gravity: A review}},}\ }}\href {\doibase
  10.1142/S0218271817300208} {\bibfield  {journal} {\bibinfo  {journal} {Int.
  J. Mod. Phys.}\ }\textbf {\bibinfo {volume} {D26}},\ \bibinfo {pages}
  {1730020} (\bibinfo {year} {2017})}\BibitemShut {NoStop}%
\bibitem [{\citenamefont {Li}\ \emph {et~al.}(2015)\citenamefont {Li},
  \citenamefont {Modesto},\ and\ \citenamefont {Rachwał}}]{Li:2015bqa}%
  \BibitemOpen
  \bibfield  {author} {\bibinfo {author} {\bibfnamefont {Y.-D.}\ \bibnamefont
  {Li}}, \bibinfo {author} {\bibfnamefont {L.}~\bibnamefont {Modesto}}, \ and\
  \bibinfo {author} {\bibfnamefont {L.}~\bibnamefont {Rachwał}},\ }\bibfield
  {title} {\emph {\enquote {\bibinfo {title} {{Exact solutions and spacetime
  singularities in nonlocal gravity}},}\ }}\href {\doibase
  10.1007/JHEP12(2015)173} {\bibfield  {journal} {\bibinfo  {journal} {JHEP}\
  }\textbf {\bibinfo {volume} {12}},\ \bibinfo {pages} {173} (\bibinfo {year}
  {2015})},\ \Eprint {http://arxiv.org/abs/1506.08619} {arXiv:1506.08619
  [hep-th]} \BibitemShut {NoStop}%
\bibitem [{\citenamefont {Modesto}\ \emph
  {et~al.}(2017{\natexlab{a}})\citenamefont {Modesto}, \citenamefont
  {Rachwal},\ and\ \citenamefont {Shapiro}}]{Modesto:2017hzl}%
  \BibitemOpen
  \bibfield  {author} {\bibinfo {author} {\bibfnamefont {L.}~\bibnamefont
  {Modesto}}, \bibinfo {author} {\bibfnamefont {L.}~\bibnamefont {Rachwal}}, \
  and\ \bibinfo {author} {\bibfnamefont {I.~L.}\ \bibnamefont {Shapiro}},\
  }\bibfield  {title} {\emph {\enquote {\bibinfo {title} {{Renormalization
  group in super-renormalizable quantum gravity}},}\ }}\href@noop {} {\
  (\bibinfo {year} {2017}{\natexlab{a}})},\ \Eprint
  {http://arxiv.org/abs/1704.03988} {arXiv:1704.03988 [hep-th]} \BibitemShut
  {NoStop}%
\bibitem [{\citenamefont {Modesto}\ \emph
  {et~al.}(2017{\natexlab{b}})\citenamefont {Modesto}, \citenamefont {Myung},\
  and\ \citenamefont {Yi}}]{Modesto:2017ycz}%
  \BibitemOpen
  \bibfield  {author} {\bibinfo {author} {\bibfnamefont {L.}~\bibnamefont
  {Modesto}}, \bibinfo {author} {\bibfnamefont {Y.~S.}\ \bibnamefont {Myung}},
  \ and\ \bibinfo {author} {\bibfnamefont {S.-H.}\ \bibnamefont {Yi}},\
  }\bibfield  {title} {\emph {\enquote {\bibinfo {title} {{Universality of the
  Unruh effect}},}\ }}\href@noop {} {\  (\bibinfo {year}
  {2017}{\natexlab{b}})},\ \Eprint {http://arxiv.org/abs/1710.04367}
  {arXiv:1710.04367 [gr-qc]} \BibitemShut {NoStop}%
\bibitem [{\citenamefont {Cornell}\ \emph {et~al.}(2017)\citenamefont
  {Cornell}, \citenamefont {Harmsen}, \citenamefont {Lambiase},\ and\
  \citenamefont {Mazumdar}}]{Cornell:2017irh}%
  \BibitemOpen
  \bibfield  {author} {\bibinfo {author} {\bibfnamefont {A.~S.}\ \bibnamefont
  {Cornell}}, \bibinfo {author} {\bibfnamefont {G.}~\bibnamefont {Harmsen}},
  \bibinfo {author} {\bibfnamefont {G.}~\bibnamefont {Lambiase}}, \ and\
  \bibinfo {author} {\bibfnamefont {A.}~\bibnamefont {Mazumdar}},\ }\bibfield
  {title} {\emph {\enquote {\bibinfo {title} {{Rotating metric in Non-Singular
  Infinite Derivative Theories of Gravity}},}\ }}\href@noop {} {\  (\bibinfo
  {year} {2017})},\ \Eprint {http://arxiv.org/abs/1710.02162} {arXiv:1710.02162
  [gr-qc]} \BibitemShut {NoStop}%
\bibitem [{\citenamefont {Frolov}(2015)}]{Frolov:2015bta}%
  \BibitemOpen
  \bibfield  {author} {\bibinfo {author} {\bibfnamefont {V.~P.}\ \bibnamefont
  {Frolov}},\ }\bibfield  {title} {\emph {\enquote {\bibinfo {title} {{Mass-gap
  for black hole formation in higher derivative and ghost free gravity}},}\
  }}\href {\doibase 10.1103/PhysRevLett.115.051102} {\bibfield  {journal}
  {\bibinfo  {journal} {Phys. Rev. Lett.}\ }\textbf {\bibinfo {volume} {115}},\
  \bibinfo {pages} {051102} (\bibinfo {year} {2015})},\ \Eprint
  {http://arxiv.org/abs/1505.00492} {arXiv:1505.00492 [hep-th]} \BibitemShut
  {NoStop}%
\bibitem [{\citenamefont {Frolov}\ and\ \citenamefont
  {Zelnikov}(2016{\natexlab{b}})}]{Frolov:2015usa}%
  \BibitemOpen
  \bibfield  {author} {\bibinfo {author} {\bibfnamefont {V.~P.}\ \bibnamefont
  {Frolov}}\ and\ \bibinfo {author} {\bibfnamefont {A.}~\bibnamefont
  {Zelnikov}},\ }\bibfield  {title} {\emph {\enquote {\bibinfo {title}
  {{Head-on collision of ultrarelativistic particles in ghost-free theories of
  gravity}},}\ }}\href {\doibase 10.1103/PhysRevD.93.064048} {\bibfield
  {journal} {\bibinfo  {journal} {Phys. Rev.}\ }\textbf {\bibinfo {volume}
  {D93}},\ \bibinfo {pages} {064048} (\bibinfo {year} {2016}{\natexlab{b}})},\
  \Eprint {http://arxiv.org/abs/1509.03336} {arXiv:1509.03336 [hep-th]}
  \BibitemShut {NoStop}%
\bibitem [{\citenamefont {Frolov}\ \emph {et~al.}(2015)\citenamefont {Frolov},
  \citenamefont {Zelnikov},\ and\ \citenamefont
  {de~Paula~Netto}}]{Frolov:2015bia}%
  \BibitemOpen
  \bibfield  {author} {\bibinfo {author} {\bibfnamefont {V.~P.}\ \bibnamefont
  {Frolov}}, \bibinfo {author} {\bibfnamefont {A.}~\bibnamefont {Zelnikov}}, \
  and\ \bibinfo {author} {\bibfnamefont {T.}~\bibnamefont {de~Paula~Netto}},\
  }\bibfield  {title} {\emph {\enquote {\bibinfo {title} {{Spherical collapse
  of small masses in the ghost-free gravity}},}\ }}\href {\doibase
  10.1007/JHEP06(2015)107} {\bibfield  {journal} {\bibinfo  {journal} {JHEP}\
  }\textbf {\bibinfo {volume} {06}},\ \bibinfo {pages} {107} (\bibinfo {year}
  {2015})},\ \Eprint {http://arxiv.org/abs/1504.00412} {arXiv:1504.00412
  [hep-th]} \BibitemShut {NoStop}%
\bibitem [{\citenamefont {Koshelev}\ and\ \citenamefont
  {Mazumdar}(2017)}]{Koshelev:2017bxd}%
  \BibitemOpen
  \bibfield  {author} {\bibinfo {author} {\bibfnamefont {A.~S.}\ \bibnamefont
  {Koshelev}}\ and\ \bibinfo {author} {\bibfnamefont {A.}~\bibnamefont
  {Mazumdar}},\ }\bibfield  {title} {\emph {\enquote {\bibinfo {title} {{Do
  massive compact objects without event horizon exist in infinite derivative
  gravity?}}}\ }}\href {\doibase 10.1103/PhysRevD.96.084069} {\bibfield
  {journal} {\bibinfo  {journal} {Phys. Rev.}\ }\textbf {\bibinfo {volume}
  {D96}},\ \bibinfo {pages} {084069} (\bibinfo {year} {2017})},\ \Eprint
  {http://arxiv.org/abs/1707.00273} {arXiv:1707.00273 [gr-qc]} \BibitemShut
  {NoStop}%
\bibitem [{\citenamefont {Calcagni}\ and\ \citenamefont
  {Modesto}(2017)}]{Calcagni:2017sov}%
  \BibitemOpen
  \bibfield  {author} {\bibinfo {author} {\bibfnamefont {G.}~\bibnamefont
  {Calcagni}}\ and\ \bibinfo {author} {\bibfnamefont {L.}~\bibnamefont
  {Modesto}},\ }\bibfield  {title} {\emph {\enquote {\bibinfo {title}
  {{Stability of Schwarzschild singularity in non-local gravity}},}\ }}\href
  {\doibase 10.1016/j.physletb.2017.09.018} {\bibfield  {journal} {\bibinfo
  {journal} {Phys. Lett.}\ }\textbf {\bibinfo {volume} {B773}},\ \bibinfo
  {pages} {596} (\bibinfo {year} {2017})},\ \Eprint
  {http://arxiv.org/abs/1707.01119} {arXiv:1707.01119 [gr-qc]} \BibitemShut
  {NoStop}%
\bibitem [{\citenamefont {Cardoso}\ \emph {et~al.}(2015)\citenamefont
  {Cardoso}, \citenamefont {Brito},\ and\ \citenamefont
  {Rosa}}]{Cardoso:2015zqa}%
  \BibitemOpen
  \bibfield  {author} {\bibinfo {author} {\bibfnamefont {V.}~\bibnamefont
  {Cardoso}}, \bibinfo {author} {\bibfnamefont {R.}~\bibnamefont {Brito}}, \
  and\ \bibinfo {author} {\bibfnamefont {J.~L.}\ \bibnamefont {Rosa}},\
  }\bibfield  {title} {\emph {\enquote {\bibinfo {title} {{Superradiance in
  stars}},}\ }}\href {\doibase 10.1103/PhysRevD.91.124026} {\bibfield
  {journal} {\bibinfo  {journal} {Phys. Rev.}\ }\textbf {\bibinfo {volume}
  {D91}},\ \bibinfo {pages} {124026} (\bibinfo {year} {2015})},\ \Eprint
  {http://arxiv.org/abs/1505.05509} {arXiv:1505.05509 [gr-qc]} \BibitemShut
  {NoStop}%
\bibitem [{\citenamefont {Barbier}\ \emph {et~al.}(2015)\citenamefont
  {Barbier}, \citenamefont {Beau},\ and\ \citenamefont
  {Goussev}}]{Barbier:2015}%
  \BibitemOpen
  \bibfield  {author} {\bibinfo {author} {\bibfnamefont {M.}~\bibnamefont
  {Barbier}}, \bibinfo {author} {\bibfnamefont {M.}~\bibnamefont {Beau}}, \
  and\ \bibinfo {author} {\bibfnamefont {A.}~\bibnamefont {Goussev}},\
  }\bibfield  {title} {\emph {\enquote {\bibinfo {title} {{Comparison between
  two models of absorption of matter waves by a thin time-dependent
  barrier}},}\ }}\href {\doibase 10.1103/PhysRevA.92.053630} {\bibfield
  {journal} {\bibinfo  {journal} {Phys. Rev.}\ }\textbf {\bibinfo {volume}
  {A92}},\ \bibinfo {pages} {053630} (\bibinfo {year} {2015})}\BibitemShut
  {NoStop}%
\bibitem [{\citenamefont {Garner}\ \emph {et~al.}(2017)\citenamefont {Garner},
  \citenamefont {Lakhtakia}, \citenamefont {Breakall},\ and\ \citenamefont
  {Bohren}}]{Garner:2017}%
  \BibitemOpen
  \bibfield  {author} {\bibinfo {author} {\bibfnamefont {T.~J.}\ \bibnamefont
  {Garner}}, \bibinfo {author} {\bibfnamefont {A.}~\bibnamefont {Lakhtakia}},
  \bibinfo {author} {\bibfnamefont {J.~K.}\ \bibnamefont {Breakall}}, \ and\
  \bibinfo {author} {\bibfnamefont {C.~F.}\ \bibnamefont {Bohren}},\ }\bibfield
   {title} {\emph {\enquote {\bibinfo {title} {{Lorentz Invariance of
  Absorption and Extinction Cross Sections of a Uniformly Moving Object}},}\
  }}\href {\doibase 10.1103/PhysRevA.96.053839} {\bibfield  {journal} {\bibinfo
   {journal} {Phys. Rev.}\ }\textbf {\bibinfo {volume} {A96}},\ \bibinfo
  {pages} {053839} (\bibinfo {year} {2017})}\BibitemShut {NoStop}%
\bibitem [{\citenamefont {Villavicencio}\ \emph {et~al.}(2018)\citenamefont
  {Villavicencio}, \citenamefont {Romo},\ and\ \citenamefont
  {Hernández-Maldonado}}]{Villavicencio:2018}%
  \BibitemOpen
  \bibfield  {author} {\bibinfo {author} {\bibfnamefont {J.}~\bibnamefont
  {Villavicencio}}, \bibinfo {author} {\bibfnamefont {R.}~\bibnamefont {Romo}},
  \ and\ \bibinfo {author} {\bibfnamefont {A.}~\bibnamefont
  {Hernández-Maldonado}},\ }\bibfield  {title} {\emph {\enquote {\bibinfo
  {title} {Absorption dynamics and delay time in complex potentials},}\ }}\href
  {http://stacks.iop.org/1402-4896/93/i=5/a=055201} {\bibfield  {journal}
  {\bibinfo  {journal} {Physica Scripta}\ }\textbf {\bibinfo {volume} {93}},\
  \bibinfo {pages} {055201} (\bibinfo {year} {2018})}\BibitemShut {NoStop}%
\bibitem [{\citenamefont {Olver}\ \emph {et~al.}(2010)\citenamefont {Olver},
  \citenamefont {Lozier}, \citenamefont {Boisvert},\ and\ \citenamefont
  {Clark}}]{Olver:2010}%
  \BibitemOpen
  \bibfield  {author} {\bibinfo {author} {\bibfnamefont {F.~W.}\ \bibnamefont
  {Olver}}, \bibinfo {author} {\bibfnamefont {D.~W.}\ \bibnamefont {Lozier}},
  \bibinfo {author} {\bibfnamefont {R.~F.}\ \bibnamefont {Boisvert}}, \ and\
  \bibinfo {author} {\bibfnamefont {C.~W.}\ \bibnamefont {Clark}},\ }\href@noop
  {} {\emph {\bibinfo {title} {NIST Handbook of Mathematical Functions}}},\
  \bibinfo {edition} {1st}\ ed.\ (\bibinfo  {publisher} {Cambridge University
  Press},\ \bibinfo {address} {New York, NY, USA},\ \bibinfo {year}
  {2010})\BibitemShut {NoStop}%
\bibitem [{\citenamefont {Boos}\ \emph {et~al.}(2018)\citenamefont {Boos},
  \citenamefont {Frolov},\ and\ \citenamefont {Zelnikov}}]{Boos:2018kir}%
  \BibitemOpen
  \bibfield  {author} {\bibinfo {author} {\bibfnamefont {J.}~\bibnamefont
  {Boos}}, \bibinfo {author} {\bibfnamefont {V.~P.}\ \bibnamefont {Frolov}}, \
  and\ \bibinfo {author} {\bibfnamefont {A.}~\bibnamefont {Zelnikov}},\
  }\bibfield  {title} {\emph {\enquote {\bibinfo {title} {{Quantum scattering
  on a delta potential in ghost-free theory}},}\ }}\href@noop {} {\  (\bibinfo
  {year} {2018})},\ \Eprint {http://arxiv.org/abs/1805.01875} {arXiv:1805.01875
  [hep-th]} \BibitemShut {NoStop}%
\bibitem [{\citenamefont {Buoninfante}\ \emph
  {et~al.}(2018{\natexlab{c}})\citenamefont {Buoninfante}, \citenamefont
  {Koshelev}, \citenamefont {Lambiase},\ and\ \citenamefont
  {Mazumdar}}]{Buoninfante:2018xiw}%
  \BibitemOpen
  \bibfield  {author} {\bibinfo {author} {\bibfnamefont {L.}~\bibnamefont
  {Buoninfante}}, \bibinfo {author} {\bibfnamefont {A.~S.}\ \bibnamefont
  {Koshelev}}, \bibinfo {author} {\bibfnamefont {G.}~\bibnamefont {Lambiase}},
  \ and\ \bibinfo {author} {\bibfnamefont {A.}~\bibnamefont {Mazumdar}},\
  }\bibfield  {title} {\emph {\enquote {\bibinfo {title} {{Classical properties
  of non-local, ghost- and singularity-free gravity}},}\ }}\href@noop {} {\
  (\bibinfo {year} {2018}{\natexlab{c}})},\ \Eprint
  {http://arxiv.org/abs/1802.00399} {arXiv:1802.00399 [gr-qc]} \BibitemShut
  {NoStop}%
\end{thebibliography}

%

\end{document}